\documentclass[aps,pra,twocolumn,floatfix,nofootinbib,showpacs,longbibliography]{revtex4-2}

\usepackage[utf8]{inputenc}  
\usepackage[T1]{fontenc}     
\usepackage[table]{xcolor}
\usepackage[british]{babel}  
\usepackage[sc,osf]{mathpazo}
\usepackage[scaled=0.86]{berasans}  
\usepackage[colorlinks=true, citecolor=blue, urlcolor=blue]{hyperref}  
\usepackage{graphicx} 
\usepackage[babel]{microtype}  
\usepackage{amsmath,amssymb,amsthm,bm,amsfonts,mathrsfs,bbm} 

\usepackage{xspace}  
\usepackage{pgf,tikz}
\usepackage{xcolor}
\usepackage{multirow}
\usepackage{array}
\usepackage{bigstrut}
\usepackage{braket}
\usepackage{color}
\usepackage{natbib}
\usepackage{multirow}
\usepackage{mathtools}
\usepackage{algorithm}
\usepackage{algpseudocode}
\usepackage{float}
\usepackage[caption = false]{subfig}
\usepackage{xcolor,colortbl}
\usepackage{color}
\usepackage{rotating}
\usepackage{tikz}
\usepackage{ulem}
\usetikzlibrary{quantikz}
\usepackage{diagbox}
\usepackage{soul}
\usepackage{tikz}
\usepackage{ragged2e}

\newcommand{\be}{\begin{equation}}
\newcommand{\ee}{\end{equation}}
\newcommand{\ba}{\begin{eqnarray}}
\newcommand{\ea}{\end{eqnarray}}

\newtheorem{definition}{Definition}
\newtheorem{proposition}{Proposition}

\def\>{\rangle}
\def\<{\langle}

\usepackage{centernot}
\usepackage{subfig}

\usepackage{soul}
\newcommand{\PB}[1]{\textcolor{orange}{#1}}
\newcommand{\PBs}[1]{\textcolor{orange}{\st{#1}}}
\newcommand{\PBc}[1]{\textcolor{orange}{[PB: #1]}}

\begin{document}
\title{Quantitative and Optimal Device-Independent Lower Bounds on Detection Efficiency}

\author{Arkaprabha Ghosal}

\author{Soumyadip Patra}

\author{Peter Bierhorst}

\affiliation{Department of Mathematics, University of New Orleans, New Orleans, LA 70148, USA}

\begin{abstract} 
This paper examines a quantitative and optimal lower bound on the detector efficiency in a (2,2,2) Bell experiment within a fully device-independent framework, whereby the detectors used in the experiment are uncharacterized. We provide a tight lower bound on the minimum efficiency required to observe a desired Bell-CHSH violation using the Navascu\'es-Pironio-Ac\'in (NPA) hierarchy, confirming tightness up to four decimal places with numerical optimization over explicit quantum realizations. We then introduce the effect of dark counts and demonstrate how to quantify the minimum required efficiency to observe a desired CHSH violation with an increasing dark count error. Finally, to obtain an analytical closed-form expression of the minimum efficiency, we consider the set of no-signaling behaviors that satisfy the Tsirelson bound, which  are easier to characterize than the quantum set. Using such behaviors, we find a simple closed-form expression for a lower bound on the minimum efficiency which is monotonically increasing with the CHSH violation, though the analytically obtained lower bounds are meaningfully below the numerically tight lower bound.       
\end{abstract}

\maketitle

\section{Introduction}
Quantum mechanics departs fundamentally from classical mechanics, which offers a purely deterministic description of physical phenomena. John Bell's seminal work \cite{PhysicsPhysiqueFizika.1.195} demonstrated that certain quantum correlations cannot be explained by any local hidden variable theory---a phenomenon known as Bell nonlocality. Bell nonlocality can be observed through experiments, designed to rule out classical explanations of certain quantum correlations. A commonly studied setting is the standard (2,2,2) scenario, where two spatially separated parties share an entangled quantum system, each party chooses between two measurement settings, and each measurement yields one of two possible outcomes. By repeating the experiment many times and comparing their recorded data, the parties can test whether the resulting correlations violate a Bell inequality (such as the Clauser–Horne–Shimony–Holt (CHSH) inequality \cite{PhysRevLett.23.880, PhysRevD.10.526}), a constraint satisfied by all local hidden variable theories. Any such violation provides compelling evidence for the nonlocal nature of quantum correlations.

In recent years, violation of Bell inequalities in Bell Test experiments have established quantum nonlocality as a valuable resource in quantum information processing. In particular, such violations enable device-independent protocols such as quantum key distribution \cite{ PhysRevLett.67.661, PhysRevLett.98.230501}, certification of quantum devices—including measurements and channels \cite{PhysRevLett.130.250201,Wagner2020deviceindependent}—and quantum random number expansion \cite{Pironio2010-zj}.

Any realistic Bell experiments will be subject to some degree of noise. A notable example is Bell experiments employing entangled photons \cite{PhysRevLett.115.250401}, which require the reliable detection of single photons in each trial \cite{PhysRevApplied.22.064067}. In practice the detectors in Alice’s and Bob’s laboratories are not perfect, meaning their detection efficiencies are typically less than 100~\% \cite{PhysRevA.57.3304, Zhao2019_loopholefree, PhysRevA.63.022117}. This imperfection directly impacts the ability to observe a violation of a Bell inequality. In Ref.~\cite{PhysRevA.47.R747}, P. H. Eberhard demonstrated that if Alice and Bob’s detectors have equal efficiency, then a violation of the Bell–CHSH inequality is impossible whenever the detection efficiency drops below $2/3$. Note that while this bound is most straightforwardly interpreted in terms of detector efficiency in the case of photonic experiments, the bound applies generally to all CHSH experiments as a bound on the probability that one measurement outcome is
erroneously switched to the other.

Since detection efficiency below $2/3$ makes violating a Bell inequality impossible, one can infer that conversely, observing a Bell violation ensures that the detectors must have had an efficiency exceeding $2/3$, and one does not need to make any detailed assumptions characterizing the detectors to draw this conclusion. Intuitively, larger quantitative violations should certify higher efficiencies, and this certification can be \textit{device-independent} because it depends only on the observed Bell violation. In this paper, we pursue this paradigm.  
Specifically, we assume that Alice and Bob each have uncharacterized detectors and share an unknown quantum state, and our central question is the following: 
\newline
\newline
{\it Given a certain amount of Bell inequality violation observed between Alice and Bob, can one predict the minimum detection efficiencies of their detectors in a fully device-independent manner?} 
\newline
\newline
This question is particularly relevant from an experimental perspective---if one purchases uncharacterized detectors from a vendor, then upon performing an experiment and observing a specific violation, it should be possible to immediately conclude that the detector efficiencies cannot be lower than a certain threshold. This protocol can also serve as a benchmark for ensuring that a quantum network is operating at a certain level of quality. We note that the above problem is the inverse to a routine familiar to experimentalists trying to implement Bell experiments without postselection \cite{PhysRevLett.111.130406, Giustina2013-je, PhysRevLett.115.250401, PhysRevLett.115.250402}, who will endeavor to maximize Bell violation based on detectors with characterized efficiency; our findings will have some implications for this task as well. 

A preliminary  version of the above highlighted problem was previously analyzed in Ref.~\cite{PhysRevLett.118.260401}. In that study, Alice and Bob were assumed to have identical, uncharacterized single-photon detectors. Each party performed dichotomic quantum measurements, with the possibility of having more than two measurement settings. The central question they addressed was: {\it if a Bell violation of any nonzero magnitude} occurs, what is the minimum threshold detector efficiency required? In the (2,2,2) Bell scenario, their predicted threshold coincides with the well-known Eberhard bound that is $2/3$. Importantly, their work did not analyze how the required efficiency depends on the amount of violation without introducing additional device-dependent assumptions, such as an assumption that one detector is perfect or that its efficiency is well characterized. Instead, higher lower bounds on efficiency were obtained in Ref.~\cite{PhysRevLett.118.260401} by adding measurement settings to find inequalities whose violation (by any nonzero amount, no matter how small) required higher minimum efficiency than Eberhard's; however, in practice adding measurement settings can be experimentally difficult. The efficiency bounds in Ref.~\cite{PhysRevLett.118.260401} were derived using the level-two Navascu\'es-Pironio-Ac\'in (NPA) hierarchy \cite{Navascués_2008, RevModPhys.86.419}, but Ref.~\cite{PhysRevLett.118.260401} did not establish whether a concrete bipartite quantum state and measurement settings can actually realize a given violation with the minimal efficiency predicted up to certain numerical accuracy, thus leaving open the question of whether the newly derived bounds were optimal (aside from the specific case of the Eberhard bound of 2/3, known to be optimal from previous work using other methods \cite{PhysRevA.63.022117}). 

In this work, we take an approach that enables us to determine the lower bound on detector efficiency in the symmetric setup required to obtain a Bell violation of any given magnitude without relying on any auxiliary assumptions. First, using numerical analysis, we minimize the detector efficiency over the set of quantum achievable probability distributions corresponding to a given amount of Eberhard violation. Our numerical results provide an upper bound on the minimum detector efficiency that could be certified from a given observed violation. For the same given Eberhard violation, we then minimize the detector efficiency using convex optimization over the set of correlations that can be realized within the level-two NPA hierarchy. The resulting optimal efficiency values serve as lower bounds on the true minimum efficiency achievable over the full set of quantum realizations. We observe that the minimum efficiency values that arise through the numerical minimization agree up to four decimal places with the minimum efficiency values arising through the level-two NPA hierarchy, and are thus optimal to this level of precision.

 We emphasize that the above interpretation of the detector efficiency bound relies on a linear noise model, as also considered in Ref.~\cite{PhysRevLett.118.260401}. In this model, an imperfect detector has the possibility of failing to show a ``{\it click}'' even if a quantum particle is there. The interpretation of this linear noise model as detector inefficiency is a form of device-dependence. However,  this can be viewed more abstractly as a probabilistic transformation from an ideal underlying behavior to the observed behavior where there is a certain chance that a ``click'' outcome is transformed into a ``no-click'' outcome. Such an effective transformation may originate from detector inefficiency itself, or from other stages such as a bit flip (from ``click'' to ``no-click'') in computer-based data registration or post-processing. 
While the interpretation of the efficiency parameter depends on the devices actually present, the lower bound on an abstract probability of one outcome being transformed into another can still be considered device independent at the level of the observed Bell statistics since the underlying devices are still treated as black boxes.

We also go beyond Ref.~\cite{PhysRevLett.118.260401} in considering the effect of dark counts in the experiment, where the detectors make a "{\it false click}" while the expected photon actually has not arrived at the lab. We define a dark count error as the probability of a "no detection" outcome to be changed to a "detection" outcome in a single experimental trial; this is the opposite effect of detector inefficiency. For a given nonzero dark count error, we observe that the efficiency values obtained through numerical minimization over quantum realizations have five decimals of agreement with the minimum values obtainable through the level-two NPA hierarchy. We also outline how this can be extended to a fully general class of linear noise models where efficiencies and dark count rates can depend on party and/or setting.

Our numerical explorations do not yield a closed-form expression for the minimum detector efficiency as an explicit function of the observed Eberhard violation. The existence of such an expression can be interesting from an analytical perspective. We perform an analytical treatment by considering the family of no-signaling behaviors that satisfy the Tsirelson bound, which is a looser superset of the collection of quantum behaviors when compared to the NPA hierarchy but easier to characterize, and find a closed form expression for a lower bound on the minimum detector efficiency. We observe that this expression is a simple and a monotonically increasing function of the observed Eberhard violation. However, we observe that these analytically obtained values are notably lower than the tight lower bound on detector efficiency obtained numerically through the NPA hierarchy.

\section{Background}

\subsection{Device-independent assumptions and Bell CHSH inequality}

 We are interested in the (2,2,2) Bell experiment where there are two spatially separated parties $A$ and $B$, they each receive one of two classical inputs, $x \in \lbrace x_0,x_1 \rbrace$ for $A$ and $y\in \lbrace y_0, y_1\rbrace $ for $B$, and they obtain two classical outputs $a\in \lbrace +1,-1 \rbrace$ and $b \in \lbrace +1,-1\rbrace$. In a device-independent scenario we have some restrictions which are imposed on the local parties, Alice and Bob \cite{Supic2020selftestingof}. Specifically, they are assumed to be at distant locations such that there is no communication between them during an experimental trial, which in the most ideal experimental paradigm can be enforced by ensuring their respective measurement events are space-like separated. Alice and Bob's outcome probabilities are assumed to be compatible with a quantum mechanical model, but no further assumptions on the states and measurements are made beyond those imposed by the spatial separation---namely, they may share a bipartite state but can perform only local quantum measurements on their respective portions of it. One also assumes the inputs $x$ and $y$ are freely chosen, and in particular not correlated with the measured quantum state. Under these device-independent assumptions, one can make inferences about the underlying state---such as, in the most basic case, the presence of entanglement-based on observed violations of Bell inequalities. Under these assumptions, we will consider a linear noise model that transforms an ideal behavior to an observed behavior without imposing any restrictions on the Hilbert-space dimension, the shared state, or the local measurement operators.
 Such model is chosen because it gives a simple operational description of the detection process: the underlying ideal behavior is first generated, and them some process probabilistically modifies the recorded outcome. It allows the relevant noise parameter to be interpreted as an effective detector efficiency and makes the connection with the standard detector-efficiency problem in Bell tests, though in practice other processes can lead to probabilistic modification of measurement outcomes from an ideal underlying behavior. A model-dependent interpretation as detector inefficiency does not invalidate the device-independent nature of the Bell analysis itself.

\subsubsection{Behaviors and probability constraints}
Considering the above assumptions for the (2,2,2) scenario, a behavior $\mathbf{P}$ is known as a set of sixteen non-negative joint conditional probabilities $\lbrace P(ab|xy)\rbrace$ i.e., $\mathbf{P}\in \mathbb{R}^{16}$. Any behavior must obey the normalization condition for each setting, $x\in \lbrace x_0, x_1 \rbrace$ and $y\in \lbrace y_0, y_1 \rbrace$, which can be described as 
\begin{align}\label{eq:Normalisation}
    \sum_{a,b} P(ab|xy) =1  \quad \quad \quad \forall x,y.
\end{align}
We say that a behavior realizes a no-signaling correlation if it can be described by the following definition.

\begin{definition} \label{def1}
A behavior $\mathbf{P}$ is {\bf no-signaling} if it satisfies the following constraints, known as the no-signaling (NS) conditions:
\begin{align}
    {\bf NS:} \quad &\sum_{b}P(ab|xy) = \sum_{b} P(ab|xy') =P(a|x) \quad \quad \forall x,y,y'\nonumber \\
    & \sum_{a} P(ab|xy) = \sum_{a} P(ab | x'y) = P(b|y) \quad \quad \forall x,x',y, \label{NS}
\end{align}
which imply that Alice cannot influence Bob's outcome and vice versa.
\end{definition}

So in a (2,2,2) scenario, we define the set of all no-signaling correlation as $\mathcal{NS}$ and the total number of constraints which are imposed on a no-signaling probability distribution is $12$ (four normalization conditions and eight no-signaling conditions). 

\vskip 0.5cm
\paragraph{Correlators and local marginals of a no-signaling behavior:} 

In the (2,2,2) scenario, we can parametrize the set of all no-signaling behavior in terms of the quantities, $\langle \hat{A}_x\rangle $, $\hat{B}_y$ and $\langle \hat{A}_x \hat{B}_y\rangle$. For $a,b\in\{+1,-1\}$, we define a correlator as the following:
\begin{align}
    \langle \hat{A}_x\hat{B}_y\rangle = \sum_{a,b} abP(ab|xy) \quad \quad \forall x,y \label{corr}
\end{align}
and we define the marginal expectation values $\langle \hat{A}_x\rangle $ and $\hat{B}_y$ as 
\begin{align}
   & \langle \hat{A}_x\rangle = \sum_{a} aP(a|x) \quad \quad \forall x\nonumber \\
   & \langle \hat{B}_y\rangle = \sum_{b} bP(b|y) \quad \quad \forall y. \label{marginal}
\end{align}
Hence, considering Eq.~\eqref{eq:Normalisation}--\eqref{marginal}, every probability distribution $P(ab|xy)$ that realizes a no-signaling correlation can be expressed as 
\begin{align}
    P(ab|xy) = \dfrac{1}{4} \left[ 1 + a\langle \hat{A}_x\rangle + b\langle \hat{B}_y\rangle + ab\langle \hat{A}_x \hat{B}_y\rangle\right] \label{probcorr}
\end{align}

\vskip 0.3cm 
\paragraph{Quantum observables:}
In the (2,2,2) scenario, a quantum observable is a Hermitian operator that has only two eigenvalues $+1$ and $-1$ and can be expressed in terms of two orthogonal projectors as 
\[
\hat{O} = \Pi_+ - \Pi_-,
\]
where $\Pi_{\pm}$ are two projectors or projective measurement locally acting on a quantum state such that $\Pi_{\pm}\Pi_{\pm} = \Pi_{\pm}$ and $\Pi_+ \Pi_- =0 $. So we define Alice's observable for setting $x$ as $\hat{A}_x$ and we define Bob's observable as $\hat{B}_y$ such that 
\begin{align}
\hat{A}_x = \Pi^x_+ - \Pi^x_-, \quad \quad  \hat{B}_y = \Pi^y_+ -  \Pi^y_-  \label{qobservable}
\end{align}
\begin{definition}
   In a (2,2,2) scenario, a behavior $\mathbf{P}$ realizes a bipartite quantum correlation if there exists a pure bipartite quantum state $\ket{\psi}_{AB}$ in a Hilbert space $\mathcal{H}$, a pair of two outcome observables $\lbrace \hat{A}_{x_0}, ~\hat{A}_{x_1}\rbrace$ for Alice, and a pair of two outcome observables $\lbrace \hat{B}_{y_0},~ \hat{B}_{y_1}\rbrace$ for Bob, such that for all $x\in \lbrace x_0, x_1\rbrace$ and $y \in \lbrace y_0, y_1 \rbrace$ 
   \[
   P(ab|xy) = \langle \Psi | \Pi^x_a \otimes \Pi_b^y \ket{\Psi},  \quad a,b\in \lbrace -1, +1\rbrace,
   \]
   where the set of measurement operators satisfy the following properties: 
   \begin{enumerate}
       \item $\Pi^x_a ~\Pi^x_{a'} = \delta_{aa'}~\Pi^x_a$, \quad and \quad $\Pi^y_b ~\Pi^y_{b'} = \delta_{bb'}~\Pi^y_b$ 
       \item $(\Pi_a^x)^{\dagger} = \Pi_a^x$, and $(\Pi_b^y)^{\dagger} = \Pi_b^y$
       \item $\sum_{a} \Pi_a^x = \mathbb{I}$, \quad and \quad $\sum_{b} \Pi_b^y = \mathbb{I}$
   \end{enumerate} \label{def2}
\end{definition}
We define the set of all quantum realizable behaviors as $\mathcal{Q}$. The set $\mathcal{Q}$ forms a strict subset of $\mathcal{NS}$ as there exists no-signaling behavior which cannot be explained as {\bf Definition} \ref{def2}. For example, the well known $PR$ box correlation \cite{PhysRevLett.95.140401} is a no-signaling correlation but cannot be realized using quantum theory.

\subsection{CHSH inequality}
In a (2,2,2) scenario, the CHSH inequality is defined as 
\begin{align}
    &|\beta_{CHSH}| \leq 2, \quad s.t. \label{CHSH} \nonumber \\
    & \beta_{CHSH} = \langle \hat{A}_{x_0}\hat{B}_{y_0}\rangle + \langle \hat{A}_{x_0}\hat{B}_{y_1}\rangle + \langle \hat{A}_{x_1}\hat{B}_{y_0}\rangle - \langle \hat{A}_{x_1}\hat{B}_{y_1}\rangle \nonumber \\
\end{align}
Any behavior $\mathbf{P}$ which can be explained through a {\it local hidden variable} (LHV) model \cite{PhysRevLett.23.880} cannot violate the above inequality.  If a behavior is quantum realizable, i.e., if $\mathbf{P}\in \mathcal{Q}$, then the CHSH inequality can be violated but the maximum CHSH violation cannot surpass the value $2\sqrt{2}$, also known as the Tsirelson bound \cite{Cirelson1980}. On the other hand, if $\mathbf{P} \in \mathcal{NS}$, then the largest violation that can be obtained is $4$ which is the algebraic maximum of the CHSH inequality.

\subsection{Imperfect detection and Eberhard Inequality}

We consider a particular physical model to motivate our problem, but the results will apply to any Bell experiment employing any physical system. Let us assume a quantum source produces a pair of entangled photons in some pure quantum state, say $\ket{\psi}_{AB}$. After that each photon travels through some optical fiber and finally Alice and Bob use polarizers and single photon detectors to obtain measurement outcomes, which may violate the CHSH inequality by a certain amount. 
Let us suppose Alice's detector has the efficiency $\eta_A$ and Bob's detector has the efficiency $\eta_B$. Here, efficiency $\eta$ of a detector means the probability that the detector makes a "{\it click}" whenever a quantum particle reaches the detector. The efficiencies are defined in such a way that the following transformation relation holds for each of the detectors: 
\[
\begin{tikzpicture}[baseline=(plus.base),>=stealth]
  \node (plus) at (0,0) {$+$};

  \node (plus2) at (2,0.8) {$+$};

  \node (zero) at (2,-0.8) {$0$};

  \draw[->] (plus) -- (plus2) node[midway,above] {$\eta_A$};
  \draw[->] (plus) -- (zero) node[midway,below] {$1-\eta_A$};

  \node (plusB) at (3,0) {$+$};
  \node (plus2B) at (5,0.8) {$+$};
  \node (zeroB) at (5,-0.8) {$0$};

  \draw[->] (plusB) -- (plus2B) node[midway,above] {$\eta_B$};
  \draw[->] (plusB) -- (zeroB) node[midway,below] {$1-\eta_B$};
\end{tikzpicture}
\]
In the above diagram, we define the symbol $``+"$ as a ``click'' of the detector that occurs when a photon successfully arrives with a certain polarization state, whereas $``0"$ implies a ``no click'' event. Note that other factors beyond the detectors, such as loss in fiber, can contribute to a probability of a ``click'' outcome being switched to ``no click'', and so later inferences about feasible values of the $\eta$ parameters should be interpreted as a lower bound on overall \textit{system} efficiency for detection. Furthermore, in a completely abstract analysis of a Bell experiment, the efficiency parameter can be interpreted in the context of a probability that one measurement outcome is switched to the other, which will be universally interpretable in any Bell experiment regardless of the physical system (though this probability could be small or even negligible depending on the implementation). The sole defining characteristic of this noise model is just that it is \textit{linear}, in the sense that the probability of outcome ``$+$'' is switched to ``$0$'' is obtained by applying the constant multiplicative factor $(1-\eta)$.

Continuing with the example of the photonic experiment, in the case of a perfect CHSH experiment with 100~\% efficiency, Alice and Bob can use a polarizer followed by a solitary photon detector in their respective labs, without any ambiguity on how to interpret ``$0$'' outcomes: Alice randomly selects the polarization axis between the settings ${x_0, x_1}$, while Bob randomly selects the polarization axis between ${y_0, y_1}$. In a single experimental trial, if either Alice or Bob obtains a “{\it click},” it implies that a photon has arrived with a certain state of polarization, corresponding to the outcome $+1$. On the other hand, if there is no click from one or both detectors, they can infer that a photon did arrive but with an orthogonal state of polarization, and was absorbed/deflected by the polarizer. Thus, without any ambiguity, they would associate a “{\it no click}” event with the outcome $-1$. However, as discussed earlier, in the case of an imperfect CHSH experiment, a “{\it no click}” event can arise for another reason, as  sometimes the detector makes an error and registers an outcome that should have resulted in ``$+$'' as ``$0$''; in other cases because a ``{\it no click}” may still imply that a photon arrived but with an orthogonal polarization state. Since these are experimentally indistinguishable, we group all such possibilities together and denote them collectively as ``$0$''.



Based on the above setup, P.~H.~Eberhard introduced an inequality in Ref.~\cite{PhysRevA.47.R747} that does not include the joint outcome ``$00$'', so it appears better suited to imperfect setups, though it is equivalent to the CHSH inequality (for no-signaling distributions). So if a no-signaling distribution violates the CHSH inequality then it also violates the Eberhard inequality and the converse is also true.  From now on, we will consider the Eberhard inequality violation for certifying the detection efficiency, since it is a little easier to apply inefficiency transformations as it comprises fewer terms than the CHSH inequality. In terms of probabilities, the Eberhard inequality can be described as follows:

\begin{align}
    &\mathcal{E} = P(++|x_0y_0) - P(+0|x_0y_1) - P(0+|x_1y_0) \nonumber \\
    &- P(++|x_1y_1) \leq 0 \label{Eberhard}
\end{align}

\paragraph{Equivalence between Eberhard and CHSH violation:} In the (2,2,2) scenario, if we consider the no-signaling set $(\mathcal{NS})$, where every behavior $\mathbf{P}$ can be described as Eq.$~$(\ref{probcorr}), one can analytically show that the Eberhard inequality $\mathcal{E}$ in Eq.$~$(\ref{Eberhard}) and the CHSH inequality $\beta_{CHSH}$ in Eq.$~$(\ref{CHSH}) are linearly related as follows: 
\begin{align}
    \mathcal{E} = \dfrac{\beta_{CHSH}}{4} - \dfrac{1}{2}. \label{equivalence}
\end{align}
The analytical proof of the above equality is given in appendix~\ref{appB}.

\section{Minimizing detector efficiency for a given amount of Eberhard violation}\label{bigsec3}
\subsection{Transformation of a behavior due to imperfect detection} \label{sec2}
In the (2,2,2) scenario, we assume that there exists a behavior $\mathbf{P}= \lbrace P(ab|xy)\rbrace$ before it is affected by the detectors that demonstrates an initial Eberhard violation of an amount $\mathcal{E}_{in}>0$. So we say that  
\begin{align}
     &P(++|x_0y_0) - P(+0|x_0y_1) - P(0+|x_1y_0) \nonumber \\
    &- P(++|x_1y_1) = \mathcal{E}_{in} 
\end{align}
After that the imperfect detectors with efficiencies $\eta_A$ and $\eta_B$ transform the initial set of probabilities $\lbrace P(ab|xy)\rbrace$ to another set of probabilities, say $\lbrace Q(ab|xy)\rbrace$, resulting another behavior $\mathbf{Q}$. The transformations of the outcomes $a,b \in \lbrace +, 0 \rbrace$ and their respective probabilities can be schematically described as follows:  
\begin{small}
\[
\begin{tikzpicture}[baseline=(plus.base),>=stealth]
  \node (pp) at (0,0) {$++$};

  \node (pp2) at (4,2.4) {$++$};

  \node (pz2)  at (4,0.8)   {$+0$};

  \node (zz1)   at (4, -2.0) {$00$};

  \node (zp2)  at (4, -0.8) {$0+$};
 
  \draw[->] (pp) -- (pp2) node[midway,above, sloped] {$\eta_A~\eta_B$};
  \draw[->] (pp) -- (pz2) node[midway,above, sloped] {$\eta_A~(1-\eta_B)$};
  \draw[->] (pp) -- (zz1) node[midway,below, sloped ] {$(1-\eta_A)~(1-\eta_B)$};
  \draw[->] (pp) -- (zp2) node[midway,above, sloped] {$(1-\eta_A)~\eta_B$};
\end{tikzpicture}
\]
\[
\begin{tikzpicture}
    
  \node (zp3)  at (0, 0) {$0+$};

  \node (zp4)  at (3,0.8) {$0+$};

  \node (zz4)  at (3,-0.8) {$00$};

   \node (pz) at (5, 0) {$+0$};

  \node (pz3) at (8,0.8) {$+0$};

  \node (zz2) at (8, -0.8) {$00$};

   \draw[->] (zp3) -- (zp4) node[midway,above, sloped] {$\eta_B$};
  \draw[->] (zp3) -- (zz4) node[midway,below, sloped] {$1-\eta_B$};
  \draw[->] (pz) -- (pz3) node[midway,above, sloped] {$\eta_A$};
  \draw[->] (pz) -- (zz2) node[midway,below, sloped] {$1-\eta_A$};

\end{tikzpicture}
\]
\end{small}

So according to the above transformation, one can write the Eberhard violation of the final behavior $\mathbf{Q}$ as 
\begin{align}
    & Q(++|x_0y_0)-Q(+0|x_0y_1)-Q(0+|x_1y_0) \nonumber \\
    &-Q(++|x_1y_1) \nonumber \\
    \nonumber \\
    & = \eta_A~\eta_B ~P(++|x_0y_0)  \nonumber \\
    &- \left[ \eta_A~P(+0|x_0y_1)+ \eta_A~(1-\eta_B)~P(++|x_0y_1)\right] \nonumber \\
    &- \left[ \eta_B ~P(0+|x_1y_0) + (1-\eta_A)~\eta_B~P(++|x_1y_0)\right] \nonumber \\
    & - \eta_A~\eta_B~ P(++|x_1y_1) \nonumber \\
    \nonumber \\
    & = \left(P_{++|x_0y_0}-P_{++|x_1y_1}+P_{++|x_0y_1}+P_{++|x_1y_0} \right)~\eta_A~\eta_B \nonumber \\
    &- \left(P_{+0|x_0y_1}+P_{++|x_0y_1}\right)~\eta_A -\left(P_{0+|x_1y_0}+P_{++|x_1y_0} \right)~\eta_B \nonumber \\
    &= \mathcal{K}~\eta_A~\eta_B - \mathcal{L}'~\eta_A - \mathcal{L}''~\eta_B, \label{transformation}
\end{align}
where we define $P_{ab|xy}$ as shorthand for the joint probability $P(ab|xy)$ and we define
\begin{align}
    &\mathcal{K} = P_{++|x_0y_0}-P_{++|x_1y_1}+P_{++|x_0y_1}+P_{++|x_1y_0}, \nonumber \\
    &\mathcal{L}'= P_{+0|x_0y_1}+P_{++|x_0y_1}, ~~\mathcal{L}''=P_{0+|x_1y_0}+P_{++|x_1y_0}. \label{Ebquant}
\end{align}
Thus if the Eberhard violation of the final behavior is $\mathcal{E}$, then we can say that observed Eberhard violation between Alice and Bob can be expressed as 
\begin{align}
    \mathcal{E} = \mathcal{K}~\eta_A~\eta_B - \mathcal{L}'~\eta_A - \mathcal{L}''~\eta_B \label{obEb}
\end{align}
and for the symmetric case when we assume that Alice and Bob's detectors are identical i.e., $\eta_A = \eta_B =\eta$, we have 
\begin{align}
    \mathcal{E} = \mathcal{K}~\eta^2 - \mathcal{L}~\eta, \label{SymEb}
\end{align}
where $\mathcal{L} = \mathcal{L}' + \mathcal{L}''$.

\subsection{Parametrization of the quantum achievable behaviors for the (2,2,2) setting} \label{para}

In the previous subsection, we described how a no-signaling behavior is modified or transformed under imperfect detection. Our goal is to determine the critical (minimum) detector efficiency required to observe a given amount of Eberhard violation. As a first step, we examine the range of efficiency values that can be achieved with a quantum model. So we want to carry out a numerical exploration over quantum realizations, which involve a bipartite entangled state and local projective measurements performed by Alice and Bob. While our numerical explorations were not guaranteed to be optimal, we will later show that the numerically obtained minimum efficiency values match the true minimum efficiency up to four decimal places,  using convex optimization at the second level of the NPA hierarchy \cite{Navascués_2008} as we will discuss in section \ref{NPA}.

For the numerical analysis, we adopt a parametrization of quantum realizations, where Alice and Bob share a pure two-qubit entangled state
\begin{align}
    \ket{\Psi}_{AB} = \cos{\theta}~\ket{00}+\sin{\theta}~\ket{11},\quad ~~ 0< \theta \leq \dfrac{\pi}{2}, \label{state}
\end{align}
and Alice, Bob perform projective measurements $\Pi^x_{a}$ for setting $x \in \lbrace x_0,x_1 \rbrace$ with outcome $a\in \lbrace +, 0 \rbrace$, and $\Pi^y_{b}$ for setting $y \in \lbrace y_0, y_1 \rbrace$ with outcome $b\in \lbrace +, 0 \rbrace$, such that
\begin{align}
    &\Pi_{+}^x =\dfrac{1}{2}(\mathbb{I}_2 + \sigma\mathbf{ \cdot \hat{n}_x}) \quad \textnormal{and} \quad  \Pi_{0}^x =\dfrac{1}{2}(\mathbb{I}_2 - \sigma\mathbf{ \cdot \hat{n}_x}),  \nonumber \\
    &\Pi_{+}^y =\dfrac{1}{2}(\mathbb{I}_2 + \sigma\mathbf{ \cdot \hat{m}_y}) \quad \textnormal{and} \quad  \Pi_{0}^y =\dfrac{1}{2}(\mathbb{I}_2 - \sigma\mathbf{ \cdot \hat{m}_y}), \label{proj}
\end{align}
where $\sigma\equiv(\sigma_X,\sigma_Y,\sigma_Z)$ is the vector of $2\times 2$ Pauli matrices, and $\mathbf{\hat{n}_x},\mathbf{\hat{n}_y}\in\mathbb{R}^3$ represent the measurement directions for Alice and Bob for settings $x$ and $y$, which are restricted in the $X$-$Z$ plane as 
\begin{align}
& \hat{n}_{x_0}  = \lbrace \sin{\theta_{a_0},~0,~\cos{\theta_{a_0}}}\rbrace, \quad  0 \leq \theta_{a_0} \leq 2~\pi \nonumber \\
& \hat{n}_{x_1}  = \lbrace \sin{\theta_{a_1},~0,~\cos{\theta_{a_1}}}\rbrace, \quad  0 \leq \theta_{a_1} \leq 2~\pi \label{setting_for_alice}
\end{align}
for Alice and
\begin{align}
& \hat{m}_{y_0} = \lbrace \sin{\theta_{b_0},~0,~\cos{\theta_{b_0}}}\rbrace, \quad  0 \leq \theta_{b_0} \leq 2~\pi \nonumber \\
& \hat{m}_{y_1}  = \lbrace \sin{\theta_{b_1},~0,~\cos{\theta_{b_1}}}\rbrace, \quad  0 \leq \theta_{b_1} \leq 2~\pi \label{setting_for_bob}
\end{align} 
for Bob. From Eq.$~$(\ref{qobservable}) and using Eq.$~$(\ref{proj}) - (\ref{setting_for_bob}) we can write the expressions of the local observables for Alice and Bob as 
\begin{small}
\begin{align}
    &\hat{A}_{x_0} = \cos{\theta_{a_0}}~\sigma_Z + \sin{\theta_{a_0}}~\sigma_X,  \quad \hat{A}_{x_1} =  \cos{\theta_{a_1}}~\sigma_Z + \sin{\theta_{a_1}}~\sigma_X,  \nonumber \\
     &\hat{B}_{y_0} = \cos{\theta_{b_0}}~\sigma_Z + \sin{\theta_{b_0}}~\sigma_X,  \quad  
    \hat{B}_{y_1} =  \cos{\theta_{b_1}}~\sigma_Z + \sin{\theta_{b_1}}~\sigma_X.  \label{parametrization}
\end{align}
\end{small}
Now using Eq.$~$(\ref{state}) and Eq.$~$(\ref{proj})--(\ref{setting_for_bob}) we can express the joint probabilities $P(ab|xy)$ as 
\begin{align}\label{born}
    P(ab|xy) = \bra{\Psi} \Pi_a^x \otimes \Pi_b^y \ket{\Psi},
\end{align}
which can be expressed as explicit functions of the five independent parameters $\theta,~ \theta_{a_0},~\theta_{a_1},~ \theta_{b_0},~\theta_{b_1}$ respectively and these functional forms are given in appendix~\ref{appA}. It is known from Ref.~\cite{PhysRevA.108.012212} and Ref.~\cite{BarizienBancal2025} that, in the (2,2,2) scenario, all extremal quantum realizations can be fully characterized by the parametrization we adopt in Eq.$~$(\ref{state})--(\ref{setting_for_bob}) (although, conversely, this parametrization does not identify only the extremal points as it contains some non-extremal ones too).

With this, one can calculate the four probability terms $P(++|x_0y_0)$, $P(+0|x_0y_1)$, $P(0+|x_1y_0)$ and $P(++|x_1y_1)$ using Eq.$~$(\ref{born}) to obtain the Eberhard quantity $\mathcal{E}$ described in Eq.$~$(\ref{Eberhard}). If we find that $\mathcal{E}>0$ for any quantum realization given in Eq.$~$(\ref{parametrization}), then we can also claim that such a realization also violates the CHSH inequality, hence, it is nonlocal.

We have discussed earlier that the Eberhard quantity in Eq.$~$(\ref{Eberhard}) is linearly related to the CHSH quantity, which is given in Eq.$~$(\ref{equivalence}). It is known that the maximum quantum violation of the CHSH quantity $\beta_{CHSH}$ is $2\sqrt{2}$, which is the Tsirelson bound \cite{Cirelson1980}. Therefore, one can conclude from Eq.$~$(\ref{equivalence}) that the maximum quantum violation of the Eberhard quantity is 
\begin{align}\label{Tbound}
    \mathcal{E}_{Q_{max}} = \dfrac{2\sqrt{2}}{4} -\dfrac{1}{2} \simeq 0.207107.
\end{align}

According to the transformation in Eq.~(\ref{Ebquant}), the above violation requires perfect detector efficiency $(\eta = 1)$, which in turn leads us to examine how smaller violations of Eberhard’s inequality can certify lower efficiencies.

\subsubsection{Performing the numerical minimization }
Now we focus on our central question. Let us consider that there is an observed Eberhard violation between Alice and Bob, say $\mathcal{E}_{obs} >0$, which can be expressed in terms of a quantum achievable behavior $\lbrace Q(ab|xy) \rbrace$ as  
\begin{align}
    & Q(++|x_0y_0) - Q(+0|x_0y_1) - Q(0+|x_1y_0) \nonumber \\
    & - Q(++|x_1y_1) = \mathcal{E}_{obs},
\end{align}
where $\mathcal{E}_{obs} \in \left( 0, \frac{\sqrt{2}-1}{2} \right]$. Now, following the transformation rule given in Eq.$~$(\ref{transformation}) and also from Eq.$~$(\ref{SymEb}), we can write the above equality condition in terms of an initial behavior $\lbrace P(ab|xy)\rbrace$ as 
\begin{align}
    \mathcal{K}~\eta^2 - \mathcal{L}~ \eta = \mathcal{E}_{obs},
\end{align}
where we consider both detectors have an efficiency $\eta$ and the quantities $\mathcal{K}$ and $\mathcal{L}$ are linear combinations of joint probabilities described in Eq.$~$(\ref{Ebquant}) and Eq.$~$(\ref{SymEb}). 

We propose the following optimization problem,
\begin{align}
    &minimize \quad \eta \nonumber \\
    & s.t. \quad \mathcal{K}~\eta^2 - \mathcal{L}~\eta = \mathcal{E}_{obs}, \nonumber \\
    & ~\quad \quad \mathbf{P} \in \mathcal{Q}, \nonumber \\
    & ~~\quad ~~0\leq \theta \leq \dfrac{\pi}{2}, ~~0 \leq \theta_{a_0}, \theta_{a_1}, \theta_{b_0}, \theta_{b_1} \leq 2~\pi. \label{numopt}
\end{align}

Now, depending on the parametrization given in Eq.$~$(\ref{parametrization}), and then from Eq.$~$(\ref{setting_for_alice})--(\ref{setting_for_bob}) one can evaluate the joint probabilities $P(ab|xy)$ as nonlinear expressions constructed from trigonometric functions of the variables $\lbrace \theta,\theta_{a_0}, \theta_{a_1}, \theta_{b_0}, \theta_{b_1}\rbrace$. The explicit functional forms of six probability terms, which are used to evaluate $\mathcal{K}$ and $\mathcal{L}$, are given in the appendix \ref{appA}. The numerical optimization in Eq.$~$(\ref{numopt}) is consequently a nonlinear optimization problem for which an analytical solution is not apparent. To numerically address the optimization problem in Eq.~(\ref{numopt}), we employ an effective method based on Algorithm~\ref{alg1}, while noting that this Algorithm is not guaranteed to converge to the global optimum. 

\begin{algorithm}[H]
\caption{Algorithm for numerical minimization using bisection method}
\begin{algorithmic}
     \State Fix $l = \tfrac{2}{3}$, $u = 1$
   \State Fix tolerance, $\epsilon$ 
   \State Define $\mathcal{K}$, $\mathcal{L}$
   \State Define $f(\eta) = \mathcal{K}~ \eta^2 - \mathcal{L}~ \eta$
   \State Define bounds: $\theta \in [0, \frac{\pi}{2}]~~ and~~ \theta_{a_0}, \theta_{a_1}, \theta_{b_0}, \theta_{b_1} \in [0, 2\pi]$
   \State Fix initial guess 
   \\
   \State Fix the value $\mathcal{E}_{obs}$ \quad $s.t.$ \quad $0< \mathcal{E}_{obs} \leq \dfrac{\sqrt{2}-1}{2}$
   \State \textbf{repeat} 
   \State Fix $\eta = \dfrac{u + l}{2}$
   \State $g(\eta)=$ Maximize$(f,~bounds,~initial~ guess)$ 
   \\
   \State \textbf{if} \quad $g(\eta)\geq \mathcal{E}_{obs}$ \quad \textbf{then} \quad $u = \eta$; \quad \textbf{else} \quad $l = \eta$
   \\
   \State \textbf{until} \quad $u - l \leq\epsilon$
   \State \textbf{output} \quad $\eta$ 
\end{algorithmic}
\label{alg1}
\end{algorithm}

If one successfully runs the above algorithm on a computer, one should obtain an efficiency value for a fixed amount of Eberhard violation $\mathcal{E}_{obs}$. Then one can repeat the same algorithm for a range of $\mathcal{E}_{obs}$ values to obtain the trend of the numerically minimized efficiency values with increasing amount of Eberhard violation. Although it may at first seem that the Algorithm \ref{alg1} takes a different approach than the proposed optimization in Eq.$~$(\ref{numopt}), both are actually equivalent. In particular, the first constraint in Eq.$~$(\ref{numopt}) requires equality with $\mathcal{E}_{obs}$, Algorithm~\ref{alg1} instead checks whether the function $f(\eta)$ is greater than $\mathcal{E}_{obs}$ (the motivation for this modification being an observation that the inequality constraint appeared a little more robust in implementing numerical optimizations). We show in appendix~\ref{appC} that for a given value of $\eta$, if there exists a quantum realization $\mathbf{P}$ that satisfies $f(\eta)> \mathcal{E}_{obs}$, then there also exists another quantum realization $\mathbf{P}'$ for which the equality $f'(\eta) = \mathcal{E}_{obs}$ is satisfied. This indicates that the formulation of Algorithm~\ref{alg1}  is equivalent with the optimization problem in Eq.$~$(\ref{numopt}) employing the equality constraint.

Algorithm \ref{alg1} represents a nested numerical optimization, where for a given value of $\mathcal{E}_{obs}\in(0, 0.2071]$, we first fix $\eta$ as the bisection of two extremal values, $u=1$ and $l=\frac{2}{3}$. Then we maximize the function $f= \mathcal{K}~\eta^2 - \mathcal{L}~\eta$ over the variables parameterizing the extremal points of $\mathcal Q$. If the maximum value is not less than $\mathcal{E}_{obs}$, we reassign the value $u \rightarrow \eta$, otherwise we reassign $l \rightarrow \eta$ and we again run the maximization until the difference $u-l$ converges to a fixed value near zero (this is the ``tolerance'' parameter in Algorithm \ref{alg1}).  In this way, the Algorithm converges to a low $\eta$ value for which the observed Eberhard quantity $\mathcal{E}_{obs}$ is still obtainable. This bisection approach allows for a simpler function $g$ to be numerically optimized (where $\eta$ is constant for each iteration), compared to direct optimization of $f$ (where $\eta$ is an additional variable). This enabled better performance with the numerical solver we implemented in {\it Python SciPy} \cite{2020SciPy-NMeth}, which minimizes a multivariable function using the Sequential Least Squares Programming (\href{https://docs.scipy.org/doc/scipy/reference/optimize.minimize-slsqp.html}{SLSQP}) method. For the numerical analysis, the bisection search is terminated once the separation between the two bisection endpoints is smaller than the chosen tolerance \(10^{-8}\).

We note the soundness of the above bisection procedure requires a minor technical observation as follows. We show in appendix~\ref{appC} that when we fix a non-negative value of the efficiency, say $\eta_0$ such that $f(\eta_0)\geq \mathcal{E}_{obs}$ for a feasible realization of $\mathcal K$ and $\mathcal L$, then any $\eta \geq \eta_0$ also satisfies the condition $f(\eta)\geq \mathcal{E}_{obs}$ for that same realization. Consequently, for a given value of $\mathcal{E}_{obs}$, the set $\mathcal S$ of $\eta$ values for which there is a quantum realization ensuring satisfaction of the inequality constraint is upper-closed. For this reason, there should exist a behavior, say $\mathbf{P}'\in \mathcal{Q}$ and a feasible efficiency value $\eta_{min}$ such that any $\eta < \eta_{min}$ cannot be a member of the set, while every $\eta > \eta_{min}$ is a member of the set.
This justifies the use of bisection search in Algorithm \ref{alg1} to find the minimum feasible: the computer searches for a lower bound of the set $\mathcal{S}$ while changing $\eta$ value at every iteration based on whether the previous $\eta$ was a member of the set $\mathcal{S}$ or not. However, it is not guaranteed that the search finds the true minimum efficiency $\eta_{min}$. This is because the search is based on an optimization of a function---that is, the function $g$ in Algorithm \ref{alg1}---that, while somewhat simplified as a result of the bisection approach, still does not fit into a category of optimization problem for which there are solvers known to provably converge to a global optimum (such as linear, or convex).

\paragraph{Using the seesaw optimization technique:}
We note that the angular parametrization used in Algorithm~\ref{alg1} is a convenient way to exploit the structure of the \((2,2,2)\) qubit realization. Nevertheless, the same problem can also be approached from an operator-level perspective. This is because 
a quantum realizable probability $P(ab|xy)$ is linearly related with a bipartite density matrix $\rho_{AB}$, and a set of local POVM elements \(\{M_a^x\}\) (for Alice) and \(\{N_b^y\}\) (for Bob), which are generalization of Eq.$~$(\ref{proj}). So for a fixed value of \(\eta\), one may therefore optimize over \(\rho_{AB}\) and local POVM elements \(\{M_a^x\}\) and \(\{N_b^y\}\) while fixing a local dimension $d\geq 2$.

In this operator formulation, the fixed-\(\eta\) optimization is not jointly convex in \(\rho_{AB}\), \(\{M_a^x\}\), and \(\{N_b^y\}\). However, if two of these three blocks are held fixed, then the optimization over the remaining block is an SDP. This leads to a standard seesaw-type procedure: one cyclically optimizes over first Bob's POVMs, then Alice's POVMs, and then the shared state, fixing the previously-optimized quantity at its optimal value when moving on to the next, and cycling through all three optimizations multiple times. Such a method can be useful for constructing feasible quantum strategies  \cite{Pal_2010, Colomer2022threenumerical} and hence for obtaining numerical upper bounds on the required detector efficiency. The important limitation is that the overall fixed-\(\eta\) problem remains nonconvex. Therefore, although each block update is an SDP, the full seesaw procedure is not guaranteed to converge to the global optimum~\cite{Pal_2010}.

 As an independent numerical check, we also implemented this seesaw-type optimization with a bisection search over \(\eta\). For each fixed value of \(\eta\), we cyclically optimize over the shared two-qubit state and the local POVM effects using SDP subproblems. For a few representative values of the observed Eberhard violation, the feasible efficiency values obtained from the seesaw procedure agree with those obtained from Algorithm~\ref{alg1} up to approximately five decimal places.

Our primary computational method in Algorithm~\ref{alg1} involving direct nonlinear parametrization is reasonable because extremal quantum behaviors in the $(2,2,2)$ scenario can be represented using two-qubit pure states and projective measurements in a real plane (due to Jordan's lemma)---a structural reduction which could also be used to restrict the see-saw method to qubits and projective measurements. However, the parametrization of Algorithm~\ref{alg1} makes this reduction explicit at the level of the optimization variables, and one can work directly with a small number of real parameters which makes multi-start searches and gradient- or Hessian-based local optimization routines transparent and computationally lightweight, rather than alternating over matrix-valued SDP subproblems.

\begin{figure}[!htb]
    \centering
    \includegraphics[width=1.0\linewidth]{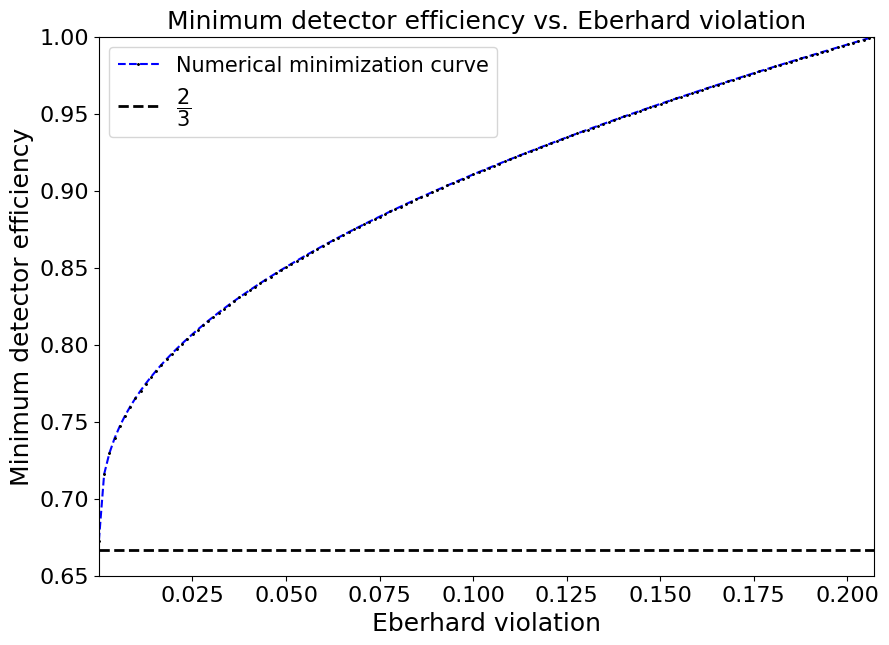}
    \caption{A plot of the numerically minimized detector efficiency $\eta^{qr}_{\mathcal{E}_{obs}}$ in the symmetric (2,2,2) setting with respect to the observed Eberhard violation $\mathcal{E}_{obs}$. The black dotted horizontal line represents the critical threshold value $\frac{2}{3}$ below which no Eberhard violation can be observed.} 
    \label{numericalplot}
\end{figure}

\paragraph{Plotting the numerical curve:} After running Algorithm \ref{alg1}, we can obtain a collection of efficiency values for the entire range of Eberhard violation, noting we do not have an analytical argument that the optimization converges to a global minimum. Thus we refer to the result of Algorithm~\ref{alg1} not as $\eta_{min}$ but as $\eta^{qr}_{\mathcal{E}_{obs}}$ with $qr$ standing for ``quantum realizable''. We plot such efficiency values with respect to an increasing amount of Eberhard violation $\mathcal{E}_{obs}$ as shown in Fig. \ref{numericalplot}.

\begin{table*}[!htb]
\caption{Detector efficiency value $\eta^{qr}_{\mathcal{E}_{obs}}$ obtained via numerical minimization for a given observed Eberhard violation $\mathcal{E}_{obs}$, yielding a quantum realization using Eq. \eqref{state}--\eqref{setting_for_bob}}
\centering
\renewcommand{\arraystretch}{1.2}
\begin{minipage}{0.7\textwidth}
\centering
\begin{ruledtabular}
\begin{tabular}{ccccccc}
$\eta^{qr}_{\mathcal{E}_{obs}}$ & $\theta$  & $\theta_{a_0}$ & $\theta_{a_1}$  &  $\theta_{b_0}$ & $\theta_{b_1}$ &  $\mathcal{E}_{obs}$ \\
\hline
$0.753774$ & $0.312188$ & $3.319115$ & $2.183795$ & $2.964065$ & $4.099364 $ & $0.006951$  \\
$0.759750$ & $0.329442$ & $3.331007$ & $2.166080$ & $2.952179$ & $4.117109$ & $0.008341$ \\
$0.765153$ & $0.344683$ & $3.341458$ & $2.151299$ & $2.941729$ & $4.131895$ & $0.009731$ \\
$0.770113$ & $0.358385$ & $3.350792$ & $2.138660$ & $2.932397$ & $ 4.144526 $ & $0.011121$ \\
$0.774718$ & $0.370863$ & $3.359227$ & $2.127669$ & $2.923957 $ & $4.155512$ & $0.012511$ \\
$0.800761 $ & $0.437283$ & $3.402773$ & $2.076648$ & $2.880402$ & $4.206528$ & $0.022241$ \\
$0.815486$ & $0.471930$ & $3.424243$ & $2.054491$ & $2.858944$ & $4.228692$ & $0.029190$ \\
$0.842109$ & $0.529795$ & $3.457401$ & $2.023453$ & $2.825780$ & $4.259724$ & $0.044480$ \\
$0.856492$ & $0.558735$ & $3.472417$ & $ 2.010502$ & $2.810750$ & $4.272637$ & $0.054210$ \\
$0.871331$ & $0.587057$ & $3.4859080$ & $1.999436$ & $2.797250 $ & $4.283721$ & $0.065330$
\end{tabular}
\end{ruledtabular}
\end{minipage}
\end{table*}

\subsection{Finding the critical detector efficiency using NPA hierarchy } \label{NPA}

Up to this point, we have shown that the numerical minimization over quantum realizations, based on our parametrization given in Eq.$~$(\ref{parametrization}), yields increasing efficiency values \begin{small}$(\eta^{qr}_{\mathcal{E}_{obs}})$ \end{small} as the Eberhard violation increases. However, as mentioned earlier it is difficult to establish analytically whether such efficiency values represent the  minimum efficiency values over all quantum realizations $\mathcal{Q}$. 

In the following, we present a method to determine a numerical lower bound on the minimum detector efficiency $\eta^{min}_{\mathcal{E}_{obs}}$ over all quantum realizations consistent with a given Eberhard violation $\mathcal{E}_{obs}$. To obtain the lower bounds, we relax the set of quantum correlations ($\mathcal{Q}$) to a larger set of correlations $(\mathcal{Q}_2)$, characterized at the second level of the NPA hierarchy \cite{Navascués_2008}. With such a relaxation technique, we can reformulate the numerical minimization problem in Eq.$~$(\ref{numopt}) as a convex optimization problem, which yields a lower bound \begin{small}$\eta^{npa}_{\mathcal{E}_{obs}}$ \end{small} on the minimum detector efficiency $\eta^{min}_{\mathcal{E}_{obs}}$. 
Conversely, the \begin{small}$\eta^{qr}_{\mathcal{E}_{obs}}$ \end{small} values obtained through Algorithm~\ref{alg1} imply an upper bound on 
$\eta^{min}_{\mathcal{E}_{obs}}$. We show that \begin{small}$\eta^{qr}_{\mathcal{E}_{obs}}$ \end{small} values agree up to four decimal places with the \begin{small}$\eta^{npa}_{\mathcal{E}_{obs}}$ \end{small} values. Such observation signifies that up to $4$ decimal places we certainly know the tight lower bound of the detector efficiency $\eta^{min}_{\mathcal{E}_{obs}}$.

We briefly review  the Navascu\'es, Pironio, Ac\'in (NPA) hierarchy, introduced in Ref. \cite{PhysRevLett.98.010401, Navascués_2008}, which provides a converging sequence, or a hierarchy of semi-definite programming (SDP) relaxations that approximate the set of quantum-achievable behaviors from outside. Let us assume that in a (2,2,2) scenario, there is a behavior $\mathbf{P}=\lbrace P(ab|xy) \rbrace$ and someone wants to certify whether or not it is quantum realizable. According to the Definition~\ref{def2}, $\mathbf{P}$ is quantum realizable if there exists a bipartite pure quantum state $\ket{\psi}_{AB}$ and local projective measurements $\Pi_{a}^x$ and $\Pi_{b}^y$ for Alice and Bob, each with two measurement settings, $x \in \lbrace x_0, x_1 \rbrace$, $y \in \lbrace y_0, y_1 \rbrace$ and two measurement outcomes $a,b$ for every $x,y$, such that 
\[
P(ab|xy) = \bra{\Psi} \Pi_a^x \otimes \Pi_b^y \ket{\Psi}.
\]
Generally, such a certification task is difficult. In this regard, the NPA hierarchy provides a series of necessary criterion to determine whether $\mathbf{P}$ belongs to nested no-signaling sets $\lbrace\mathcal{Q}_k \rbrace_{k=1}^{\infty}$, where each $\mathcal{Q}_k$ (for $k>1$) is a strict subset of $\mathcal{Q}_{k-1}$. In the limit $k\rightarrow \infty$ the true quantum set $\mathcal{Q}$ is contained in $\mathcal{Q}_{\infty}$. While testing whether $\mathbf{P}\in \mathcal{Q}_k$, the NPA hierarchy provides a semi-definite programming (SDP) for each level $k$, where the objective is to find a moment matrix $\Gamma_k$ which is a semidefinite matrix $(\Gamma_k \succeq 0)$ whose elements can be expressed as 
\begin{align}
    \Gamma_k^{(i,j)} = \bra{\psi} \mathcal{O}_i^{\dagger}\mathcal{O}_j \ket{\psi}, \label{momentelement}
 \end{align}
where $\mathcal{S}_k = \lbrace \mathcal{O}_i\rbrace$ denotes a collection of operators $\mathcal{O}_i$ which are linear combinations of products of projective measurement operators, where the length of the product is at most $k$. Therefore, every $\mathcal{O}_i$ is not necessarily Hermitian and the operator $\mathcal{O}_0$ for every $k$ is an identity operator. For this reason, we put the constraint that \[\Gamma_k^{(0,0)} =1. \]

In the SDP, there should be a set of constraints imposed on the elements of $\Gamma_k$ such as \cite{PhysRevLett.98.010401, Bowles2020boundingsetsof, Navascués_2008}
\begin{align}
    \sum_{m,n} g^{(i)}_{mn} ~\Gamma_k^{(n,m)} =0,  \label{cons1}
\end{align}
where $\lbrace g^{(i)}_{m,n} \rbrace$ are scalar coefficients and also 
\begin{align}
    \sum_{m',n'} f^{(j)}_{m'n'} ~\Gamma_k^{(n',m')} = \sum_{l} c_{jl} ~P_l,   \label{cons2}
\end{align}
where $c_{jl}, ~ f^{(j)}_{m'n'}\in \mathbb{R}$, and $P_l$ is the $l^{th}$ component of the behavior $\mathbf{P}$.
 For example, the constraint equality in Eq.$~$(\ref{cons1}) can capture a property of the matrix $\Gamma_k$ when multiple elements of $\Gamma_k$ are the same, for example, if $\Gamma_k^{(m,n)} = \Gamma_k^{(o,p)} = \Gamma_k^{(q, r)}$ is implied by (\ref{momentelement}), then from   Eq.$~$(\ref{cons1}) we can write,
\[\Gamma_k^{(m,n)}- \Gamma_k^{(o,p)} =0, ~~\Gamma_k^{(m,n)}- \Gamma_k^{(q,r)} =0.\]
Furthermore, \eqref{cons2} can arise in a situation where a linear combinations of a subset of $\Gamma_k$ entries represent some linear combinations of a subset of components of the behavior $\mathbf{P}$.
The constraint equalities in Eq.$~$(\ref{cons1}) and Eq.$~$(\ref{cons2}) 
can be alternatively represented with the collections of matrices $\lbrace G_i \rbrace$ and $\lbrace F_j \rbrace$ such that Eq.$~$(\ref{cons1}) and (\ref{cons2}) can be expressed as 
\[
Tr \left[ G_i ~\Gamma_k\right] =0 , \quad Tr\left[ F_j ~\Gamma_k\right] = \sum_{l} c_{jl}~ P_l \quad \quad \forall i,j.
\]

After defining the certificate matrix $\Gamma_k$ and the possible constraints it obeys, we can propose the SDP as follows \cite{PhysRevLett.98.010401, Bowles2020boundingsetsof}:
\begin{align}
    &Find \quad \Gamma_k \nonumber \\
    &s.t. \quad   \Gamma_k^{(0,0)} =1 \nonumber \\
    & \quad \quad  Tr\left[ G_i~\Gamma_k\right] =0 \quad \forall i \nonumber \\
    & \quad \quad Tr\left[ F_j ~\Gamma_k\right] = \sum_{l}c_{jl}~P_l \quad \forall j \nonumber \\
    & \quad \quad  \Gamma_k \succeq 0, \label{sdp1}
\end{align}

For the (2,2,2) setting we can write out the sequence of operators $\mathcal{S}_k$ in Eq.$~$(\ref{momentelement}) for different level-$k$ as follows:  
\begin{align}
    \mathcal{S}_1 = &\left \lbrace  \mathbb{I}, ~\hat{A}_{x_0}, ~\hat{A}_{x_1}, ~\hat{B}_{y_0},~\hat{B}_{y_1}\right \rbrace  \nonumber \\
     \mathcal{S}_{1+AB} = &\left \lbrace  \mathbb{I}, ~\hat{A}_{x_0}, \hat{A}_{x_1}, ~\hat{B}_{y_0},~\hat{B}_{y_1},~ \hat{A}_{x_0}\hat{B}_{y_0},~ \hat{A}_{x_0}\hat{B}_{y_1}, \right. \nonumber \\
    &\left.\hat{A}_{x_1}\hat{B}_{y_0},~\hat{A}_{x_1}\hat{B}_{y_1}\right \rbrace \nonumber \\
     \mathcal{S}_2 = & \left \lbrace  \mathbb{I}, ~\hat{A}_{x_0}, \hat{A}_{x_1}, ~\hat{B}_{y_0},~\hat{B}_{y_1},~\hat{A}_{x_0}\hat{A}_{x_1},~\hat{A}_{x_0}\hat{B}_{y_0} \right. \nonumber \\
    &\left.\hat{A}_{x_0}\hat{B}_{y_1},~ \hat{A}_{x_1}\hat{A}_{x_0},\hat{A}_{x_1}\hat{B}_{y_0},\hat{A}_{x_1}\hat{B}_{y_1},\hat{B}_{y_0}\hat{B}_{y_1},\hat{B}_{y_1}\hat{B}_{y_0}\right \rbrace \nonumber \\ \label{sequence}
\end{align}
where $\hat{A}_x$ and $\hat{B}_y$ are Hermitian operators which can be written as $\hat{A}_x = \Pi_x^{+} - \Pi_x^{-}$ and $\hat{B}_y = \Pi_y^{+} - \Pi_y^{-}$, where $\Pi^{\pm}_{x,y}$ are projectors. Note here the product terms like $\hat{A}_x \hat{B}_y$ can be represented as $(\hat{A}_x \otimes \mathbb{I}) ~(\mathbb{I} \otimes \hat{B}_y)$, where Alice and Bob are spatially separated and perform local measurements. So without any loss of generality we assume that the commutation relation
\[
[\hat{A}_{x},  ~\hat{B}_{y}  ] =0 
\]
holds for all $x,y$. Hence, from the sequences given in Eq.$~$(\ref{sequence}), one can construct moment matrices $\Gamma_1$, $\Gamma_{1+AB}$, $\Gamma_2$ and so on for the (2,2,2) setting using Eq.$~$(\ref{momentelement}). Note the matrix dimension for $\Gamma_1$ is $5 \times 5$, for $\Gamma_{1+AB}$ is $9 \times 9$, and for $\Gamma_2$ is $13 \times 13$. So for instance, the matrix representation of $\Gamma_1$ can be written as \cite{Navascués_2008}
\begin{align}
    \Gamma_1 = \left( 
    \begin{array}{ccccc}
        1 & \langle \hat{A}_{x_0} \rangle & \langle \hat{A}_{x_1}\rangle  & \langle \hat{B}_{y_0} \rangle & \langle \hat{B}_{y_1}\rangle  \\
        \cdot & 1 &  \langle \hat{A}_{x_0}\hat{A}_{x_1} \rangle & \langle \hat{A}_{x_0} \hat{B}_{y_0} \rangle & \langle \hat{A}_{x_0} \hat{B}_{y_1} \rangle \\
        \cdot & \cdot & 1 & \langle \hat{A}_{x_1} \hat{B}_{y_0}\rangle & \langle \hat{A}_{x_1}\hat{B}_{y_1} \rangle \\
        \cdot & \cdot & \cdot & 1 & \langle \hat{B}_{y_0}\hat{B}_{y_1} \rangle \\
        \cdot & \cdot & \cdot & \cdot & 1
    \end{array} \right).
\end{align}

\paragraph{Minimizing detector efficiency under level-two NPA hierarchy:}
Using the NPA hierarchy, we want to solve the following minimization problem: 
\begin{align}
    &\min_{\mathcal{K},\mathcal{L}} \quad \eta \nonumber \\
    &s.t. \quad \mathcal{K}~\eta^2 - \mathcal{L}~\eta \geq \mathcal{E}, \nonumber \\
    & \quad \quad \mathcal{K} = P_{++|x_0y_0} - P_{++|x_1y_1}+P_{++|x_0y_1} + P_{++|x_1y_0}, \nonumber \\
    & \quad \quad \mathcal{L} = P_{+0|x_0y_1} + P_{++|x_0y_1}+P_{0+|x_1y_0} + P_{++|x_1y_0},\nonumber \\
    & \quad \quad \lbrace P_{ab|xy} \rbrace \in \mathcal{Q}_2. \label{opt1npa}
\end{align}

Now to solve this problem using NPA methods, we first need to represent the variables $\mathcal{K}$ and $\mathcal{L}$ in Eq.$~$(\ref{Ebquant}) in terms of the $NPA_2$ moment matrix elements. Using Eq.$~$(\ref{probcorr}) we can write the quantities $\mathcal{K}$ and $\mathcal{L}$ as 
\begin{small}
\begin{align}
    \mathcal{K} &=  \dfrac{1}{2}+\dfrac{\langle \hat{A}_{x_0} \rangle + \langle \hat{B}_{y_0} \rangle}{2} + \dfrac{\langle \hat{A}_{x_0} \hat{B}_{y_0}\rangle + \langle \hat{A}_{x_0} \hat{B}_{y_1}\rangle}{4}\nonumber  \\
    & + \dfrac{\langle \hat{A}_{x_1} \hat{B}_{y_0}\rangle - \langle \hat{A}_{x_1} \hat{B}_{y_1}\rangle}{4} \nonumber \\
    \mathcal{L} & = 1+\dfrac{\langle \hat{A}_{x_0} \rangle + \langle \hat{B}_{y_0} \rangle}{2}.
\end{align}
\end{small}
Therefore, $\mathcal{K}$ and $\mathcal{L}$ can be expressed as a linear combinations of the moment matrix $\Gamma_2$ as follows: 
\begin{small}
\begin{align}
    &\mathcal{K} = \dfrac{1}{2} + \dfrac{\Gamma_2^{(0,1)} + \Gamma_2^{(0,3)}}{2}+ \dfrac{\Gamma_2^{(0,6)} + \Gamma_2^{(0,7)}+ \Gamma_2^{(0,9)} - \Gamma_2^{(0,10)}}{4} \nonumber \\
    & \mathcal{L} =  1 + \dfrac{\Gamma_2^{(0,1)} + \Gamma_2^{(0,3)}}{2}. \label{K,L,gamma}
\end{align}
\end{small}
Note here that the above expression of $\mathcal{K,~L}$ corresponds to the constraint equality in Eq.$~$(\ref{cons2}). The reason is that $\mathcal{K,~L}$ are defined as a linear combinations of the joint probabilities, which can be also represented as a linear combinations of the elements of $\Gamma_2$.
Hence, the optimization problem proposed in Eq.$~$(\ref{opt1npa}) can be represented in terms of $\Gamma_2$ elements as 
\begin{small}
\begin{align}
    &\min_{\lbrace \Gamma_2\rbrace} \quad \eta \nonumber \\
     &s.t. \quad  \mathcal{K}~\eta^2 - \mathcal{L}~\eta \geq \mathcal{E}, \nonumber \\
     &   \Gamma_2 \succeq 0, \quad \Gamma_k^{(i,i)}=1, \quad \quad \forall i, \nonumber \\
     & \mathcal{K} = \dfrac{1}{2} + \dfrac{1}{2}\left(\Gamma_2^{(0,1)} + \Gamma_2^{(0,3)}\right) + \dfrac{1}{4}\left(\Gamma_2^{(0,6)} + \Gamma_2^{(0,7)}+ \Gamma_2^{(0,9)} - \Gamma_2^{(0,10)}\right), \nonumber \\
    &  \mathcal{L} = 1 + \dfrac{1}{2}\left(\Gamma_2^{(0,1)} + \Gamma_2^{(0,3)}\right), \nonumber \\ 
    &  \sum_{k,l} c_{k,l}^{(i)} ~\Gamma_{2}^{(k,l)} =0   \quad \quad c_{k,l}^{(i)} \in \mathbb{R}, ~~ \forall ~i\in [1,40], \label{opt2}
    \end{align}
\end{small}    
where apart from the diagonal entries, there are a total of forty necessary constraint equalities on the $\Gamma_2$ matrix elements in Eq.~\eqref{opt2} which are explicitly given below:
\begin{widetext}
    \begin{small}
    \begin{align}
     &  \Gamma_2^{(0,1)} = \Gamma_2^{(2,8)} = \Gamma_2^{(3,6)} = \Gamma_2^{(4,7)},  \Gamma_2^{(0,2)} = \Gamma_2^{(1,5)} = \Gamma_2^{(3,9)} = \Gamma_2^{(4,10)}, \quad \Gamma_2^{(0,3)} = \Gamma_2^{(1,6)} = \Gamma_2^{(2,9)} = \Gamma_2^{(4,12)}, \nonumber \\
    &\Gamma_2^{(0,4)} = \Gamma_2^{(1,7)} = \Gamma_2^{(2,10)} = \Gamma_2^{(3,11)}, \quad  \Gamma_2^{(0,5)} = \Gamma_2^{(1,2)} = \Gamma_2^{(6,9)} = \Gamma_2^{(7,10)}, \quad \Gamma_2^{(0,6)} = \Gamma_2^{(1,3)} = \Gamma_2^{(8,9)} = \Gamma_2^{(7,12)}, \nonumber \\
    & \Gamma_2^{(0,7)} = \Gamma_2^{(1,4)} = \Gamma_2^{(6,11)} = \Gamma_2^{(8,10)}, \quad \Gamma_2^{(0,9)} = \Gamma_2^{(2,3)} = \Gamma_2^{(5,6)} = \Gamma_2^{(10,12)}, \quad \Gamma_2^{(0,10)} = \Gamma_2^{(2,4)} = \Gamma_2^{(5,7)} = \Gamma_2^{(9,11)}, \nonumber \\
    &\Gamma_2^{(0,11)} = \Gamma_2^{(3,4)} = \Gamma_2^{(6,7)} = \Gamma_2^{(9,10)}, \quad \Gamma_2^{(1,9)} = \Gamma_2^{(3,5)}, ~~ \Gamma_2^{(1,10)} = \Gamma_2^{(4,5)},~~ \Gamma_2^{(1,11)} = \Gamma_2^{(3,7)},~~\Gamma_2^{(1,12)} = \Gamma_2^{(4,6)}, \nonumber \\
    & \Gamma_2^{(2,6)} = \Gamma_2^{(3,8)}, ~~\Gamma_2^{(2,7)} = \Gamma_2^{(4,8)},~~ \Gamma_2^{(2,11)} = \Gamma_2^{(3,10)},~~\Gamma_2^{(2,12)} = \Gamma_2^{(4,9)},~~\Gamma_2^{(6,10)} = \Gamma_2^{(8,11)},~~ \Gamma_2^{(7,9)} = \Gamma_2^{(8,12)}. 
    \label{opt2npa}
\end{align}
\end{small}
\end{widetext}
However, the above optimization problem does not qualify as a semidefinite program (SDP) if $\eta$ is treated as a free variable, since this renders the first constraint nonlinear. We can convert the above problem to a valid SDP, where we do not treat $\eta$ as a free varible, rather we fix $\eta$ to a value within its feasible range, bounded by the extremal points $\eta_{\max}=1$ and $\eta_{\min}=\tfrac{2}{3}$. Then for each fixed $\eta$, we solve the following SDP, while expressing $\mathcal{K}$, $\mathcal{L}$ as given in Eq.$~$(\ref{K,L,gamma}) and enforcing the $NPA_2$ constraints specified in Eq.~(\ref{opt2npa}):  
\begin{small}
\begin{align}
    & Given \quad \eta \quad s.t. \quad  \dfrac{2}{3}< \eta \leq 1 \nonumber \\
    &\max_{\lbrace \Gamma_2^{(i,j)} \rbrace} \quad f_{cost} = \mathcal{K}~\eta^2 - \mathcal{L}~\eta \nonumber \\
    & \quad s.t. \quad \Gamma_2 \succeq 0, \quad \Gamma_2^{(i,i)}=1 \quad \quad \forall i \nonumber \\
    & \quad \quad \mathcal{K} = \dfrac{1}{2}+ \sum_{m,n} \alpha_{m,n} ~\Gamma_2^{(m,n)} , \quad \quad \quad \alpha_{m,n}\in \mathbb{R} \nonumber \\
    & \quad \quad \mathcal{L} = 1 + \sum_{m',n'} \beta_{m',n'}~ \Gamma_{2}^{(m',n')}, \quad \quad \beta_{m',n'} \in \mathbb{R} \nonumber \\ 
    & \quad \quad  \sum_{k,l} c_{k,l}^{(i)} ~\Gamma_{2}^{(k,l)} =0   \quad \quad c_{k,l}^{(i)} \in \mathbb{R}, ~~ \forall ~i\in [1,40],\label{sdp1}
\end{align}
\end{small}
and we use this SDP for minimizing the detector efficiency while considering $NPA_2$ hierarchy. If there exists any optimal feasible solution which cannot be less than a certain violation, say any $\mathcal{E} \in (0, 0.2071]$, we modify the extremal points, $\eta_{max}$, $\eta_{min}$, accordingly so that after certain iterations their difference converges near zero. This is the same bisection method we already adopted while doing the numerical minimization in the preceding section; however, even if we fix a $\eta$ value in Algorithm \ref{alg1}, it does not become a SDP as occurs here. So we proceed with the following Algorithm:
\begin{algorithm}[H]
\caption{Algorithm for the nested SDP under level-two NPA hierarchy}
\begin{algorithmic}
     \State Fix $l = \tfrac{2}{3}$, $u = 1$
   \State Fix tolerance, $\epsilon$ 
   \State Define $\mathcal{K}$, $\mathcal{L}$ in terms of $\Gamma_2^{(i,j)}$
   \State Define $f(\eta) = \mathcal{K}~ \eta^2 - \mathcal{L}~ \eta$
   \State Define necessary constraints of $\Gamma_2$
   \\
   \State Fix the value  $\mathcal{E}_{obs}$ \quad s.t. \quad $0 < \mathcal{E}_{obs} \leq \dfrac{\sqrt{2}-1}{2}$
   \State \textbf{repeat} 
   \State Fix $\eta = \dfrac{u + l}{2}$
   \\
   \State \textbf{solve} the SDP given in Eq.$~$(\ref{sdp1})  
   \\
   \State \textbf{if} $~~$ optimal $f_{cost}\geq \mathcal{E}_{obs}$ \quad \textbf{then} \quad $u = \eta$; \quad \textbf{else} \quad $l = \eta$
   \\
   \State \textbf{until} \quad $u - l \leq\epsilon$
   \State \textbf{output} \quad $\eta$ 
\end{algorithmic}
\label{alg2}
\end{algorithm}
 In the previous section, we have argued that the bisection method is a sound approach to numerically determine the minimized efficiency $\eta$, since for a fixed violation $\mathcal{E}_{obs}$, the set of efficiency values satisfying the inequality constraint in Algorithm~\ref{alg1} forms an upper-closed set. The same upper-closed argument works as well for Algorithm~\ref{alg2}. (To elaborate, this is because the proof of upward-closure does not rely on any property of the entries of $\mathbf P$ beyond the nonnegativity of the probabilities, and nonnegativity will hold at 2nd and higher levels of the NPA hierarchy as discussed in Appendix D of~\cite{Navascués_2008}, or could also just be directly enforced by adding constraints to the SDP.) 

Since the feasible set of $\eta$ values is upper-closed, the bisection method determines the smallest efficiency $\eta_{\mathcal{E}_{obs}}^{npa}$ such that the constraint $f_{cost}(\eta_{\mathcal{E}_{obs}}^{npa})\ge\mathcal{E}_{obs}$ is satisfied. For any fixed $\eta$, the function $f_{cost}$ is linear in the components of $\mathbf P$, so the underlying optimization is convex and the solver returns a global optimum for that $\eta$. Consequently, by repeatedly applying bisection in $\eta$, a computer implementation of the above algorithm will find the global minimum efficiency $\eta_{\mathcal{E}_{obs}}^{npa}$ compatible with the observed violation $\mathcal{E}_{obs}$ over the set of correlations realizable at the level $NPA_{2}$.

The bisection step is performed over the (initial) interval $(2/3,1]$. At each value of $\eta$, the SDP gives us a binary piece of information, i.e., whether $f_{cost}^\ast(\eta)\ge\mathcal{E}_{obs}$ or $f_{cost}^\ast(\eta)<\mathcal{E}_{obs}$, thereby determining on which side of the threshold the current value of $\eta$ lies. After $r$ bisection steps, the interval length is $(1/3)2^{-r}$. To ensure this length is at most the prescribed tolerance, we require $(1/3)2^{-r}\le\epsilon$, or equivalently, $r\ge \log_{2}(1/(3\epsilon))$. Thus, the number of SDP solves required by the outer bisection loop is bounded by 
\begin{equation*}
N = \left\lceil\log_2\left(\frac{1}{3\epsilon}\right) \right\rceil.
\end{equation*}
For the tolerance $\epsilon=10^{-8}$ used in our numerical implementation, this gives $N=25$. Hence, the overhead due to the outer scalar search is only logarithmic in the requested precision. The outer-loop complexity is thus $\Theta(\log_2(1/\epsilon))$ SDP solves.

Before proceeding, we highlight here a divergence in our approach from that of the earlier work 
Ref.~\cite{PhysRevLett.118.260401}. There, the task for the NPA hierarchy in the fully device-independent analysis is limited to finding quantum realizations that enable surpassing the critical threshold of $\mathcal E = 0$ in \eqref{SymEb}, allowing a simplification of the expression for the critical threshold from $\mathcal K \eta^2 - \mathcal L \eta =0$ to $\mathcal K \eta - \mathcal L =0$, which corresponds to Eq.~(8) of \cite{PhysRevLett.118.260401}. However, for fixed positive violations---where this cancellation does not occur---the upward closure of the set of feasible $\eta$ ensures the bisection Algorithm with the NPA hierarchy is still effective for finding the critical $\eta$; consequently, we can proceed with the analysis of nonzero violations without introducing additional device-dependent assumptions. This also illustrates that conducting similar efficiency analyses for other Bell inequalities does not require the local bound to be zero. 

\begin{figure}[!htb]
    \centering
    \includegraphics[width=1.0\linewidth]{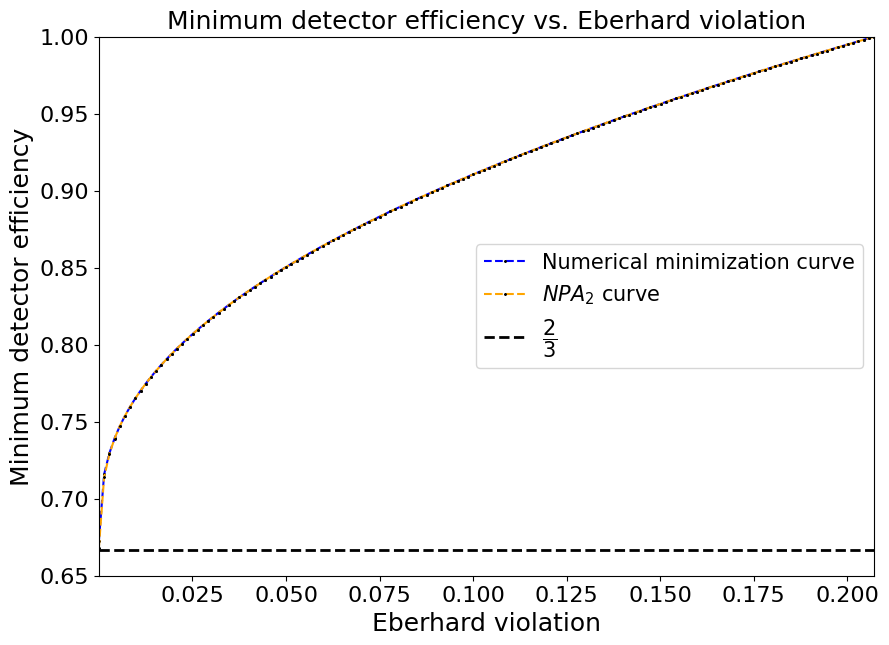}
    \caption{
    This figure shows two numerical plots of the minimum detector efficiency \(\eta\) corresponding to different observed Eberhard violation values. The blue dotted curve reproduces the numerical minimization curve from Fig.~1, obtained using Algorithm~1. The yellow dotted curve represents the minimum-efficiency values obtained using Algorithm~2, which gives a lower bound on the detector efficiency \(\eta\) over the set \(\mathcal{Q}_2\). No visible gap appears between the two plots, since the numerical values obtained from Algorithm~1 agree up to four decimal places with the corresponding values obtained under the \(NPA_2\) relaxation, as shown in Table~\ref{table2}. The bisection tolerance in Algorithms~1 and 2 is set to \(10^{-8}\). }  
    \label{npa2curve}
\end{figure}

To obtain the trend of  $\eta^{npa}_{\mathcal{E}_{obs}}$ with increasing amount of Eberhard violation, one needs to repeat Algorithm \ref{alg2} for the entire range of Eberhard violation. Below we plot $\eta^{npa}_{\mathcal{E}_{obs}}$ with respect to the Eberhard violation $\mathcal{E}_{obs}$ in Fig. \ref{npa2curve} and in the same figure we also plot the numerical curve given in Fig. \ref{numericalplot} to show a comparison between $\eta^{qr}_{\mathcal{E}_{obs}}$ and $\eta^{npa}_{\mathcal{E}_{obs}}$ values. 

\begin{table}[!htb]
\caption{Comparisons between the numerical values which are plotted in Fig.~\ref{npa2curve}}
\centering
\renewcommand{\arraystretch}{1.2}
\begin{tabular*}{0.7\columnwidth}{@{\extracolsep{\fill}}ccc@{}}
\hline\hline
\textbf{$\eta^{qr}_{\mathcal{E}_{obs}}$} & $\eta^{npa}_{\mathcal{E}_{obs}}$  & $\mathcal{E}_{obs}$ \\ 
\hline
$0.753774$ & $0.753773$ & $0.006951$  \\
$0.759750$ & $0.759749$ & $0.008341$ \\
$0.765153$ & $0.765152$ & $0.009731$  \\
$0.770113$ & $0.770112$ & $0.011121$ \\
$0.774718$ & $0.774717$ & $0.012511$ \\
$0.800761$ & $0.800760$ & $0.022241$ \\
$0.815486$ & $0.815484$ & $0.029190$ \\
$0.842109$ & $0.842107$ & $0.044480$ \\
$0.871331$ & $0.871329$ & $0.065330$ \\
\hline\hline
\end{tabular*}
\label{table2}
\end{table}

Table [\ref{table2}] shows that for any given amount of Eberhard violation $(\mathcal{E}_{obs})$, the numerically optimized efficiency $\eta^{qr}_{\mathcal{E}_{obs}}$ agrees up to four decimal places with $\eta^{npa}_{\mathcal{E}_{obs}}$, which is a lower bound on the minimum efficiency over the quantum set $\mathcal{Q}$. Here $\eta^{npa}_{\mathcal{E}_{obs}}$ is obtained by the nested convex optimization over the set $\mathcal{Q}_2 \supseteq \mathcal{Q}$, and such optimization always converges to the global optimum (on $\mathcal Q_2$). 

 In our numerical implementation of Algorithm~\ref{alg2}, the bisection search over \(\eta\) is terminated when the interval between the upper and lower bounds satisfies \(u-l\leq 10^{-8}\). This tolerance controls the resolution of the bisection search over \(\eta\). We note, however, that this bisection tolerance should not be interpreted as implying that all numerical methods must agree to eight decimal places. The final comparison between the heuristic angular optimization in Algorithm~\ref{alg1} and the NPA-based SDP method in Algorithm~\ref{alg2} is also affected by other numerical factors, including the precision of the nonlinear optimization, the SDP solver tolerances, numerical conditioning, and the fact that the two methods are not analytically guaranteed to produce exactly identical values.

In addition, the SDP comparison is performed at the second level of the NPA hierarchy, which is an outer approximation to the true quantum set. Therefore, the $\mathrm{NPA}_2$ values need not coincide exactly with the values obtained from a direct quantum realization or heuristic angular optimization. Higher levels of the NPA hierarchy may further tighten the relaxation and potentially improve the agreement. In the present work, we use the second level because it already provides a computationally tractable and numerically strong benchmark.  Therefore, although the bisection interval is refined to \(10^{-8}\), the observed agreement to four decimal places between the two independent numerical approaches should be understood as robust practical agreement---suitable for customary levels of experimental precision---rather than as proof of exact analytical coincidence.



Thus the above agreement signifies that up to four decimal places we can certainly predict the lowest possible detector efficiency (the actual minima) over the entire set $\mathcal{Q}$, when there is an observed Eberhard violation $\mathcal{E}_{obs}$. This is the best-possible device-independent bound on minimum efficiency that can be certified from a fixed Eberhard violation. We remark that our agreement results suggest that in computing the inverse problem mentioned in the introduction---optimizing Bell violation for characterized detectors, a common pre-experiment routine for photonic experiments closing the detection loophole \cite{PhysRevLett.111.130406, Giustina2013-je, PhysRevLett.115.250401, PhysRevLett.115.250402}---there can be a reasonable expectation that numerical search such as in Section \ref{para} will converge to globally optimal states, a result of interest for experimentalists performing this routine. For checking optimality, the globally optimal Bell violation achievable for a fixed value of $\eta$ can be upper bounded by the NPA hierarchy; recent results of Ref.~\cite{Gigena2025} also derive an analytic expression for this value.


\vspace{-4mm}
\section{Minimizing detector efficiency in the presence of dark counts }

Sub-unit efficiency---or a nonzero probability of turning a ``click'' event ($+$) into a ``no-click'' event ($0$)---is not the only possible error model for a realistic photon detector. In this section, we consider a more general 
scenario where Alice's and Bob's detectors make dark count errors---converting a ``no-click'' event into a ``click'' event---along with their property of making imperfect detections. 

That is, both Alice's and Bob's detectors can mistakenly click even if no particle actually arrived. So the dark count error transforms the observed output as follows: 
\[
\begin{tikzpicture}[>=stealth,scale=0.85,every node/.style={scale=0.85}]
  \node (zero) at (0,0) {$0$};

  \node (plus) at (2,0.8) {$+$};

  \node (zero2) at (2,-0.8) {$0$};

  \draw[->] (zero) -- (plus) node[midway,above, sloped] {$\xi$};
  \draw[->] (zero) -- (zero2) node[midway,below, sloped] {$1-\xi$};
\end{tikzpicture}
\]
where $\xi$ denotes the probability of a dark count error. So $\xi = 0.02$ means that there is $2\%$ chance that the detector makes a {\it false click} in a single experimental trial. This is another type of linear noise, where now the false-click rate is a linear function of the \textit{no-click} probability. Therefore, if we include the effect of dark count, the joint outcomes are transformed as follows:
\[
\begin{small}
\begin{tikzpicture}[>=stealth,scale=0.8,every node/.style={scale=0.8}]
  \node (pp) at (0,3) {$++$};

  \node (pp2) at (4,5.4) {$++$};

  \node (pz2)  at (4,3.8)   {$+0$};

  \node (zz1)   at (4, 0.8) {$00$};

  \node (zp2)  at (4, 2.2) {$0+$};

  \draw[->] (pp) -- (pp2) node[midway,above, sloped] {$\eta^2$};
  \draw[->] (pp) -- (pz2) node[midway,above, sloped] {$\eta~(1- \eta)$};
  \draw[->] (pp) -- (zz1) node[midway,below, sloped ] {$(1-\eta)^2$};
  \draw[->] (pp) -- (zp2) node[midway,above, sloped] {$(1-\eta)~\eta$};

  \node (pz3)  at (4.8, 3) {$+0$};

  \node (pz4)  at (8.8,5.4) {$+0$};

  \node (zp4)  at (8.8,3.8) {$0+$};

  \node (zz4)  at (8.8, 2.2) {$00$};

  \node (pp4) at (8.8, 0.6) {$++$};
  \node (zp) at (4.8,-2.4) {$0+$};

  \node (zp1) at (8.8, -0.2) {$0+$};

  \node (pz1) at (8.8, -1.6) {$+0$};

  \node (zz3) at (8.8, -3.4) {$00$};

  \node (pp1) at (8.8, -5.0) {$++$};

 \draw[->] (zp) -- (zp1) node[midway,above, sloped] {$\eta ~ (1-\xi)$};
  \draw[->] (zp) -- (pz1) node[midway,above, sloped] {$ \xi~(1-\eta)$};
  \draw[->] (zp) -- (zz3) node[midway,above, sloped] {$(1-\eta)~(1- \xi)$};
  \draw[->] (zp) -- (pp1) node[midway,above, sloped] {$\xi ~\eta$};
  
  \draw[->] (pz3) -- (pz4) node[midway,below, sloped] {$\eta~(1-\xi)$};
  \draw[->] (pz3) -- (zp4) node[midway,below, sloped] {$(1-\eta)~\xi$};
  \draw[->] (pz3) -- (zz4) node[midway,below, sloped ] {$(1-\eta)(1-\xi)$};
  \draw[->] (pz3) -- (pp4) node[midway,below, sloped] {$\eta~\xi$};

  \node (zz) at (0.0, -2.4) {$00$};

  \node (zz5) at (4, -0.2) {$00$};

  \node (pz5) at (4, -1.6) {$+0$};

  \node (zp5) at (4, -3.4) {$00$};

  \node (pp5) at (4, -5.0) {$++$};

  \draw[->] (zz) -- (zz5) node[midway,below, sloped] {$ (1-\xi)^2$};
  \draw[->] (zz) -- (pz5) node[midway,below, sloped] {$ \xi~(1-\xi)$};
  \draw[->] (zz) -- (zp5) node[midway,below, sloped] {$(1-\xi)~\xi$};
  \draw[->] (zz) -- (pp5) node[midway,below, sloped] {$\xi^2$};
\end{tikzpicture}
\end{small}
\]
and according the above transformation of outcomes, one can obtain a transformed noisy behavior $\mathbf{Q} =\lbrace Q(ab|xy)\rbrace$ from an initial behavior, say $\mathbf{P} =\lbrace P(ab|xy) \rbrace$. We can express the Eberhard violation of the transformed behavior $\mathbf{Q}$ as
\begin{small}
\begin{align}
    & Q(++|x_0y_0)-Q(+0|x_0y_1)-Q(0+|x_1y_0)-Q(++|x_1y_1) \nonumber \\
    & = \left[\eta^2 ~P_{++|x_0~y_0} + \xi^2 ~P_{00| x_0 ~y_0} + \xi~\eta ~P_{0+|x_0~y_0} + \xi~\eta~P_{+0|x_0~y_0}\right] \nonumber \\
    &-\left[\eta~(1-\xi)~P_{+0|x_0~y_1} + \xi~(1-\eta)~P_{0+|x_0~y_1} + \eta ~(1-\eta)~P_{++|x_0~y_1} \right. \nonumber \\
     &\left. + \xi~(1-\xi)~P_{00 | x_0~y_1} \right]  \nonumber \\
    &- \left[ \eta~(1-\xi)~ P_{0+ |x_1~y_0}+ (1-\eta)~\xi~P_{+0|x_1~y_0} + \eta~(1-\eta)~P_{++|x_1~y_0}  \right. \nonumber \\
    & \left.+ (1-\xi)~\xi~P_{00|x_1~y_0}\right] \nonumber \\
    &- \left[\eta^2 ~P_{++|x_1~y_1} + \xi^2 ~P_{00| x_1 ~y_1} + \xi~\eta ~P_{0+|x_1~y_1} + \xi~\eta~P_{+0|x_1~y_1}\right] \nonumber \\
    \nonumber \\
     &= \left( P_{++|x_0y_0} - P_{++|x_1y_1} + P_{++|x_0y_1} + P_{++|x_1y_0}\right)~ \eta^2 \nonumber \\ 
    & - \left(  P_{++|x_0y_1}  + P_{++|x_1y_0} +P_{+0|x_0y_1} + P_{0+|x_1y_0}\right) ~\eta \nonumber \\
    &+ \left( P_{0+|x_0y_0} + P_{+0|x_0y_0} + P_{0+|x_0~y_1} + P_{+0|x_1~y_0} \right. \nonumber \\
    &\left.+ P_{+0|x_0y_1} + P_{0+|x_1y_0} - P_{0+|x_1y_1} - P_{+0|x_1y_1}\right)~\xi~\eta \nonumber \\
    &- \left( P_{00|x_0y_1} +P_{00|x_1y_0} + P_{0+ |x_0~y_1} +P_{+0|x_1~y_0}\right)~\xi \nonumber \\
    & + \left( P_{00|x_0y_0} + P_{00|x_0y_1} + P_{00|x_1y_0}- P_{00|x_1y_1}\right) ~\xi^2. \label{obsEbdark}
\end{align}    
\end{small}
Therefore, if $\mathcal{E}_{obs}$ denotes the observed Eberhard violation of the behavior $\mathbf{Q}$, we can write the following expression, 

\begin{align}
    \mathcal{E}_{obs} &= Q(++|x_0y_0)-Q(+0|x_0y_1)-Q(0+|x_1y_0) \nonumber \\
    &-Q(++|x_1y_1) \nonumber \\
    & = \mathcal{K}~\eta^2 - \mathcal{L}~\eta + \mathcal{M} ~\xi~\eta - \mathcal{N}~\xi + \mathcal{O} ~\xi^2, \label{darkeq}
\end{align}
where we have 
\begin{align}
     \mathcal{K} &= P_{++|x_0y_0} - P_{++|x_1y_1} + P_{++|x_0y_1} + P_{++|x_1y_0}, \nonumber \\
    \mathcal{L} &= P_{++|x_0y_1}  + P_{++|x_1y_0} +P_{+0|x_0y_1} + P_{0+|x_1y_0} \nonumber \\
    \mathcal{M} &=  P_{0+|x_0y_0} + P_{+0|x_0y_0} +P_{0+|x_0~y_1} + P_{+0|x_1~y_0} \nonumber \\
    &+ P_{+0|x_0y_1} + P_{0+|x_1y_0} - P_{0+|x_1y_1} - P_{+0|x_1y_1}, \nonumber \\
    \mathcal{N} &=  P_{00|x_0y_1} +P_{00|x_1y_0} + P_{0+|x_0~y_1} + P_{+0|x_1~y_0}, \nonumber\\
    \mathcal{O} &= P_{00|x_0y_0} + P_{00|x_0y_1} + P_{00|x_1y_0}- P_{00|x_1y_1} \label{quantdarkcount}
\end{align}
\vskip 0.3cm
which are linear combinations of the joint probabilities and the quantities $\mathcal{K},~\mathcal{L}$ are previously defined in Eq.$~$(\ref{Ebquant}). Therefore, Eq.$~$(\ref{darkeq}) represents a nonlinear equation in terms of $\eta $ and $\xi$. 

\subsection{Minimizing \texorpdfstring{$\eta$}{eta} for a fixed value of \texorpdfstring{$\xi$}{xi}}
We now study an optimization problem by fixing the dark count error $\xi$ to a small value. Under this condition, our goal is to determine the minimum detector efficiency $\eta$ needed for a fixed amount of observed violation $\mathcal{E}_{obs}$, similar to the analysis in the previous section (\ref{para}) but this time including dark count error. The optimization problem can be proposed as follows: 
\begin{align}
    & Given \quad \quad \xi \nonumber \\
    & minimize \quad \eta \nonumber \\
    & s.t. \quad \mathcal{K}~\eta^2 - \mathcal{L}~\eta + \mathcal{M} ~\xi~\eta - \mathcal{N}~\xi + \mathcal{O} ~\xi^2 = \mathcal{E}_{obs} \nonumber \\
    & \quad \quad ~\mathbf{P} \in \mathcal{Q}
    \label{numoptdark}
\end{align}
where $\mathcal{K,L,M, N, O}$ are linear combinations of joint probabilities described in Eq.$~$(\ref{quantdarkcount}).
\vskip 0.2cm 
\paragraph{Performing numerical minimization over quantum realizations:}
We consider the parametrization given in Eq.$~$(\ref{parametrization}) to perform a numerical analysis over quantum realizations. As there are five independent parameters in Eq.$~$(\ref{parametrization}), we evaluate all sixteen joint probabilities of the behavior $\mathbf{P}$,
using Eq.$~$(\ref{setting_for_alice}) - (\ref{setting_for_bob}). The explicit expressions of these probabilities are given in appendix~\ref{appA}.

\begin{figure}[!htb]
    \centering
    \includegraphics[width=1.0\linewidth]{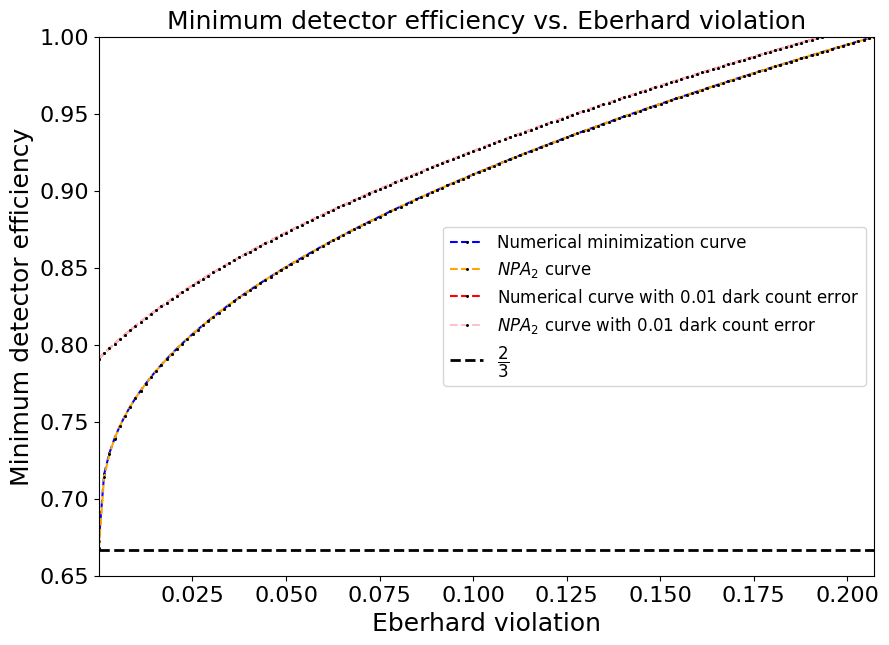}
    \caption{
   This figure shows four numerical plots of the minimum detector efficiency. The blue and yellow dotted curves reproduce, respectively, the numerical minimization curve and the \(NPA_2\)-based minimization curve from Fig.~\ref{npa2curve} for \(\xi=0\). The red and pink dotted curves show the corresponding numerical and \(NPA_2\)-based minimization curves for \(\xi=0.01\). No visible gap appears between the red and pink dotted curves; the numerical comparison is shown in Table~\ref{table2}. The bisection searches are terminated with tolerance \(10^{-8}\).} 
    \label{npa2darkc}
\end{figure}

We then evaluate the quantities $\mathcal{K}$, $\mathcal{L}$, $\mathcal{M}$, $\mathcal{N}$ and $\mathcal{O}$ using Eq.$~$(\ref{quantdarkcount}). After that we follow Algorithm~\ref{alg1} to numerically minimize $\eta$ for a given violation $\mathcal{E}_{obs}$ that satisfies the nonlinear constraint Eq.$~$(\ref{darkeq}) while fixing $\xi = 0.01$. It is important to mention here that for a fixed nonzero $\xi$, the set of efficiency values that satisfying the inequality $\mathcal{K}~\eta^2 - \mathcal{L}~\eta + \mathcal{M} ~\xi~\eta - \mathcal{N}~\xi + \mathcal{O} ~\xi^2 \geq \mathcal{E}_{obs}$ forms an upper-closed set; the proof is a straightforward generalization of the argument presented in appendix \ref{appC}. For this reason, the iterative bisection method mentioned in Algorithm~\ref{alg1} efficiently works to numerically minimize the efficiency value depending on the parametrization given in Eq.$~$(\ref{parametrization}).  We then repeat the algorithm for different amounts of violation and obtain numerically minimized efficiency values $\eta^{qr}_{\mathcal{E}_{obs}}$ for the entire range of Eberhard violation. We plot the numerical values $\eta^{qr}_{\mathcal{E}_{obs}}$ in Fig.~\ref{npa2darkc} while fixing $\xi =0$ and $\xi = 0.01$ respectively. 
\vskip 0.1cm
For $\xi= 0.01$, we find numerically that the observed Eberhard violation $(\mathcal{E}_{obs})$ in Eq.$~$(\ref{darkeq}) achievable with the parametrization given in Eq.$~$(\ref{parametrization}) cannot surpass the value $0.193201$. Otherwise, the $\eta^{qr}_{\mathcal{E}_{obs}}$ value crosses the value $1.0$, which is not a feasible solution. As mentioned earlier that Algorithm~\ref{alg1} is not a convex optimization; hence, it is not guaranteed that the numerical values represent the true minimum or not. However, we will see that these values agree up to five decimal places while we perform minimization over the behaviors realizable at the level $NPA_2$, confirming that this is the true minimum to a high degree of numerical precision.

\vskip 0.3cm
\paragraph{Performing convex optimization using $NPA$ hierarchy:}
As discussed in Section~\ref{NPA}, to obtain the true minimum value of $\eta$ we relax the quantum set $\mathcal{Q}$ to the larger set $\mathcal{Q}_2$, which is realizable at the level $NPA_2$. Accordingly, we reformulate the optimization problem in Eq.~\eqref{numoptdark} by introducing the moment matrix $\Gamma_2$ (defined in Eq.~\eqref{momentelement} and Eq.~\eqref{sequence}). In this formulation, the quantities $\mathcal{K}$, $\mathcal{L}$, $\mathcal{M}$, $\mathcal{N}$, and $\mathcal{O}$ are expressed in terms of the entries of $\Gamma_2$, in the same manner we saw previously in Eq.~\eqref{opt2}. 

After that we adopt Algorithm~\ref{alg2}, where we fix the value of $\xi = 0.01$ and run the nested convex optimization. If one can successfully run the algorithm, one can obtain a lower bound $ \eta^{npa}_{\mathcal{E}_{obs}}$ on the minimum efficiency $\eta_{\mathcal{E}_{obs}}^{min}$ over all quantum realizations, and for a given amount of Eberhard violation $\mathcal{E}_{obs}$. Finally to obtain the trend of $\eta_{\mathcal{E}_{obs}}^{npa}$ with increasing amount of $\mathcal{E}_{obs}$, one needs to repeat the algorithm for the entire range of violation. We then plot the $\eta_{\mathcal{E}_{obs}}^{npa}$ values with respect to the amount of violation in Fig.~\ref{npa2darkc} and one can see the there is no visible gap between the numerically obtained values $(\eta_{\mathcal{E}_{obs}}^{qr})$ and the true minimum values of NPA-obtained lower bounds on $\eta_{\mathcal{E}_{obs}^{min}}$, indicating tightness of the lower bound to numerical precision.

\begin{table}[!htb]
\caption{Comparison between the numerical values of $\eta$ obtained via numerical minimization and  convex optimization using $NPA_2$ which are plotted in Fig. \ref{npa2darkc} for $\xi = 0.01$.}
\centering
\renewcommand{\arraystretch}{1.2}
\begin{tabular*}{0.7\columnwidth}{@{\extracolsep{\fill}}ccc@{}}
\hline\hline
$\eta^{qr}_{\mathcal{E}_{obs}}$ & $\eta^{npa}_{\mathcal{E}_{obs}}$ & $\mathcal{E}_{obs}$ \\
  \hline
  $0.80656515$ & $0.80656469$ & $0.006951$  \\
  $0.80938961$ & $0.80938909$ & $0.008341$ \\
  $0.81213722$ & $0.81213697$ & $0.009731$  \\
  $0.81481379$ & $0.81481343$ & $0.011121$ \\
  $0.81742444$ & $0.81742399$ & $0.012511$ \\
  $0.83418241$ & $0.83418145$ & $0.022241$ \\
  $0.84488939$ & $0.84488813$ & $ 0.029190$ \\
  $0.86588749$ & $0.86588573$ & $0.044480$ \\
  $0.89065631$ & $0.89065397$ & $0.065330$ \\
  \hline\hline
\end{tabular*}
\label{table3}
\end{table}

In the above table, it can be seen that the numerically minimized efficiency values \begin{small}$(\eta^{qr}_{\mathcal{E}_{obs}})$ \end{small} agree with the efficiency values \begin{small}$(\eta^{npa}_{\mathcal{E}_{obs}})$ \end{small} up to five decimal places while we fix $\xi =0.01$ ( $1\%$ chance of dark count in the experiment), where every \begin{small}$\eta^{npa}_{\mathcal{E}_{obs}}$ \end{small}  is obtained by minimizing over the $NPA_2$ set. We also tried  different values of $\xi$ and obtained the efficiency values using Algorithm~\ref{alg1} and \ref{alg2}, and for each case we found a similar agreement between \begin{small}$\eta^{qr}_{\mathcal{E}_{obs}}$ \end{small} and \begin{small}$\eta^{npa}_{\mathcal{E}_{obs}}$ \end{small}. The observed trend of the minimum efficiency values in Fig.~\ref{npa2darkc} tells us that higher $\xi$ would lift the curve higher, and also the maximum possible value of the observed violation, corresponding to unit efficiency, would decrease as well. 
These observations comport with intuition, as a higher dark count error would introduce more noise to the system, requiring higher detector efficiency to observe an Eberhard violation.

We remark that the noisy detector model we have considered until now can be generalized further. Up to this point, we have not considered that the efficiency of the detectors could be sensitive to the measurement settings of Alice and Bob. Such a property may occur while running an actual experiment as it is conceivable that a detector could function a little differently on loss depending on the setting. In such a case, we can consider that for Alice's detector, there exist two independent efficiency parameters, $\eta_A^{x_0}$ for setting $x_0$, and $\eta_A^{x_1}$ for setting $x_1$, respectively. Similarly for Bob, we would have $\eta_{B}^{y_0}$ for setting $y_0$ and $\eta_B^{y_1}$ for setting $y_1$. Considering the dark count effect, we can also introduce two independent dark count errors, $\xi_A^{x_0}$ and $\xi_A^{x_1}$ for Alice, and similarly, $\xi_A^{x_0}$ and $\xi_A^{x_1}$ for Bob. 

Therefore, there would be a total of eight linear noise parameters arising due to imperfect detection. In such a complicated setup, one could generalize the expression of the Eberhard quantity in Eq.$~$(\ref{darkeq}) that could be highly nonlinear in terms of the eight parameters. However, as we have done in this section, we can fix any seven parameters to some positive values not exceeding one, and we may possibly obtain a tight lower bound on the uncharacterized parameter using NPA hierarchy. In this way one could partition the whole 8-dimensional space of parameters into a "feasible" and "non-feasible" region for a given value of observed Eberhard violation, with the two regions separated by a 7-dimensional boundary manifold determined by repeatedly sampling values of the fixed parameters and optimizing the free one.

\section{Finding analytical lower bound on detection efficiency from specific no-signaling distributions}

So far, we have numerically obtained a tight lower bound on the detector efficiency $\eta_{\mathcal{E}_{obs}}^{min}$ over all quantum realizations for a given Eberhard violation $\mathcal{E}_{obs}$, accurate up to four decimal places. 
However, the numerical exploration described in previous sections could not produce a closed-form expression of the minimum detector efficiency as a function of the Eberhard violation. 
In this section, we present a method for finding an analytic expression for a lower bound on detection efficiency, and compare it to the numerically tight bound $\eta_{\mathcal{E}_{obs}}^{min}$.

Our method employs a relaxation of the quantum set to no-signaling behaviors satisfying the Tsirelson bound. We start with the fact that any no-signaling behavior $P_{Q}$ violating the CHSH inequality (\ref{CHSH}) 
has the following representation \cite{Bierhorst_2016}, 
\begin{align}
    P_{Q}&= p_{PR}~PR_1 + p_1~D_1 + p_4~D_4 + p_5~D_5 + p_8~D_8 \nonumber \\
    &+ p_9 ~D_9 + p_{12}~D_{12}+ p_{14}~D_{14}+ p_{15}~D_{15}, \label{Pcorr}
\end{align} 
which denotes a convex mixture of one $PR$ box correlation and the eight local deterministic (LD) correlations that saturate the CHSH inequality, such that the $p$ coefficients are nonnegative and
\[
p_{PR}+ p_1 +p_4 +p_5 + p_8 + p_9 +p_{12}+ p_{14}+ p_{15}=1
\]
is satisfied. The index numbers of the $p_i$ and $D_i$ appearing above correspond to the enumeration of all (sixteen) local deterministic distributions in Table A2 of \cite{Bierhorst_2016}. In Eq.$~$(\ref{Pcorr}), $PR_1$ represents the one of the eight possible no-signaling PR box correlations \cite{Bierhorst_2016} which violates the CHSH inequality (\ref{CHSH})  and has the form
\begin{small}
\begin{align}
PR_1 = \begin{array}{c|c|c|c|c|}
         & ++ &+0 & 0+ & 00 \\
         \hline 
        x_0 ~y_0 & \dfrac{1}{2} & 0 &  0 & \dfrac{1}{2} \\
         & & & & \\
        \hline 
        x_0~y_1 & \dfrac{1}{2} & 0 & 0 & \dfrac{1}{2} \\
        & & & & \\
        \hline 
        x_1~y_0 & \dfrac{1}{2} & 0 & 0 & \dfrac{1}{2} \\
        & & & & \\
        \hline 
        x_1~y_1 & 0 & \dfrac{1}{2} & \dfrac{1}{2} & 0
    \end{array} \label{PR1} 
\end{align}
\end{small}
If we consider the above $PR$ box representation and we consider the representations of the eight LD correlations $(D_1, ~D_4,~D_5,~D_8,~D_9,~D_{12},~D_{14},~D_{15})$ enumerated in Ref.~\cite{Bierhorst_2016} that saturate \eqref{CHSH}, then we can explicitly write the representation of $P_Q$ in Eq.$~$(\ref{Pcorr}) as
\begin{widetext}
\begin{small}
\begin{align}
    P_{Q}= \begin{array}{c|c|c|c|c|}
         & ++ &+0 & 0+ & 00 \\
         \hline 
        x_0 ~y_0 & p_1+p_5+p_9+\dfrac{p_{PR}}{2} & p_{14} &  p_{15} & p_4 +p_8 +p_{12}+\dfrac{p_{PR}}{2} \\
         & & & & \\
        \hline 
        x_0~y_1 & p_1+p_9+p_{14}+\dfrac{p_{PR}}{2} & p_5 & p_8 & p_4+p_{12}+p_{15}+\dfrac{p_{PR}}{2} \\
        & & & & \\
        \hline 
        x_1~y_0 & p_1+p_5+p_{15}+\dfrac{p_{PR}}{2} & p_{12} & p_9 & p_4+p_8+p_{14}+\dfrac{p_{PR}}{2} \\
        & & & & \\
        \hline 
        x_1~y_1 & p_1 & p_5+p_{12}+p_{15}+\dfrac{p_{PR}}{2} & p_8+p_{9}+p_{14}+\dfrac{p_{PR}}{2} & p_4 \label{Pcorrtab}
    \end{array}
\end{align}
\end{small}
\end{widetext}
From the above representation, we can calculate the Eberhard violation of $P_Q$ using Eq.$~$(\ref{Eberhard}) as follows, 
\begin{align}
    &\mathcal{E}(P_{Q})= P_Q(++|x_0y_0)- P_Q(+0|x_0y_1)- P_Q(0+| x_1 y_0)  \nonumber \\
    &- P_Q(++|x_1y_1) \nonumber \\
    &= \dfrac{p_{PR}}{2}.\label{EbpPR}
\end{align}
An important point to notice here is that if $P_Q$ is quantum realizable, the Tsirelson bound \eqref{Tbound} must be obeyed, so the weight on $PR_1$ cannot surpass  
\[
p_{PR} \leq 2~ \mathcal{E}_{max}(P_Q) = \sqrt{2}- 1.
\]

Now, let us introduce the effect of imperfection detection without a dark count error (with an efficiency $\eta$), which allows the transformation described in Eq.$~$(\ref{transformation}) from $P_Q$ to another no-signaling behavior $P'_{Q,\eta}$.  Using Eq.$~$(\ref{transformation}), Eq.$~$(\ref{Eberhard}) and Eq.$~$(\ref{Pcorrtab}), we can express the Eberhard violation of the distribution $P'_{Q,\eta}$ as
\begin{align}
    \mathcal{E}(P_{Q},\eta) &= \left( \dfrac{3}{2}~p_{PR} + 2~(p_1+p_5+p_9)+ p_{14}+p_{15}\right)\eta^2 \nonumber \\ 
    &- \left( p_{PR} + 2~(p_1+p_5+p_9)+ p_{14}+p_{15}\right)~\eta. \label{eb1}
\end{align}
\newline
Note that the equality in Eq.$~$(\ref{eb1}) does not depend on three weights, $p_4$, $p_8$ and $p_{12}$ of the initial behavior $P_Q$.  For simplicity let us define the nonnegative quantity $\alpha = 2~(p_1+p_5+p_9)+ p_{14}+p_{15}$.
If we assume that the behavior $P'_{Q, \eta}$ has an Eberhard violation $\mathcal{E}_{obs}$ such that $0< \mathcal{E}_{obs}\leq \frac{1}{2}(\sqrt{2}-1)$, then from Eq.$~$(\ref{eb1}) the following equality will hold:
\begin{align}
    \left( \dfrac{3}{2} p_{PR} + \alpha \right) ~\eta^2 - (p_{PR} + \alpha)~\eta = \mathcal{E}_{obs}. \label{ebereq}
\end{align} 
\vskip 0.3cm 

We will now construct the most robust no-signaling behavior $P_{ns}$ which satisfies the equality condition in Eq.$~$(\ref{ebereq}) for the lowest possible detector efficiency.  Consider a no-signaling behavior comprising a convex mixture of the $PR_1$ correlation in Eq.$~$(\ref{PR1}) and three LD behaviors as follows:
\begin{align}
    &P_{ns} = q_{PR} ~PR_1 + p_4 ~D_4 + p_8~D_8 + p_{12}~D_{12}, \nonumber \\
    \text{where} \quad &q_{PR}= \sqrt{2}-1, \quad p_4 + p_8 +p_{12} = 1-q_{PR}.  \label{pns}
\end{align}

\noindent Eq.(\ref{ebereq}) reduces to the following expression for $P_{ns}$:
\begin{align}\label{quadratic}
    \mathcal{E}_{obs}=\mathcal{E}(P_{ns},\eta)  =   \dfrac{3~q_{PR}}{2}~\eta^2 - q_{PR}~\eta.
\end{align}
Above, $\mathcal{E}(P_{ns},\eta)$ is strictly increasing in $\eta$ for $\eta \in [2/3, 1]$ with a value of $0$ for $\eta=2/3$, and $q_{PR}/2 = (\sqrt{2}-1)/2$ which is the quantum maximum for $\eta = 1$. Hence for a fixed $\mathcal{E}_{obs}$ in the range $\big (0, (\sqrt{2}-1)/2\big]$ the corresponding $\eta_{ns}\in (2/3, 1]$ that yields $\mathcal{E}_{obs}$ is found by the solving \eqref{quadratic} for $\eta$. The only positive real root of the above quadratic equation has the form, 
\begin{align}
    \eta_{ns} = \dfrac{1}{3} + \dfrac{1}{3}\sqrt{1+ \dfrac{6~\mathcal{E}_{obs}}{q_{PR}}}, \quad \quad q_{PR}= \sqrt{2} - 1. \label{effns}
\end{align}
\vskip 0.2cm
Now the following proposition shows us that, given an observed Eberhard violation $\mathcal{E}_{obs}>0$, to find the minimal possible $\eta$ consistent with $\mathcal{E}_{obs}$ for any \eqref{Pcorr}-type distribution satisfying the Tsirelson bound, it is sufficient to consider the restricted family of distributions of form \eqref{pns} for which $\eta$ is given by \eqref{effns}.
\begin{proposition}
  Suppose $\mathcal{E}(\mathcal{P}_Q, ~\eta) = \mathcal{E}_{obs}>0$ holds for some no-signaling behavior $P_Q$ that violates the Eberhard inequality while satisfying the Tsireson bound. Then there exists a different no-signaling behavior $P_{ns}$ of the form \eqref{pns}, comprising a convex combination of $\sqrt 2 - 1$ weight on the the $PR_1$ box correlation and all remaining weight on (only) the three local deterministic correlations $D_4, D_8, D_{12}$, such that $\mathcal{E}(\mathcal{P}_{ns}, ~\eta_{ns}) = \mathcal{E}_{obs}$ is satisfied for some $\eta_{ns} \leq \eta $. \label{prop1}
\end{proposition}

    The proof of the above proposition is given in Appendix~\ref{appD}. The proposition implies that \eqref{effns} provides a lower bound on efficiency for all expressions of the form \eqref{Pcorr} that satisfy the Tsirelson bound, which in turn includes (strictly) all quantum behaviors violating the Eberhard inequality.  
    Below we plot the analytically obtained minimum efficiency values $(\eta_{ns})$ and the numerical lower bound on the detector efficiency over quantum realizations $(\eta_{\mathcal{E}_{obs}}^{min})$ (up to $4$ decimals) with respect to the observed violation $\mathcal{E}_{obs}$. While the analytic curve is much easier to calculate, it does fall significantly below the optimal and tight numerical lower bound.

\begin{figure}[!htb]
    \centering
\includegraphics[width=1.0\linewidth]{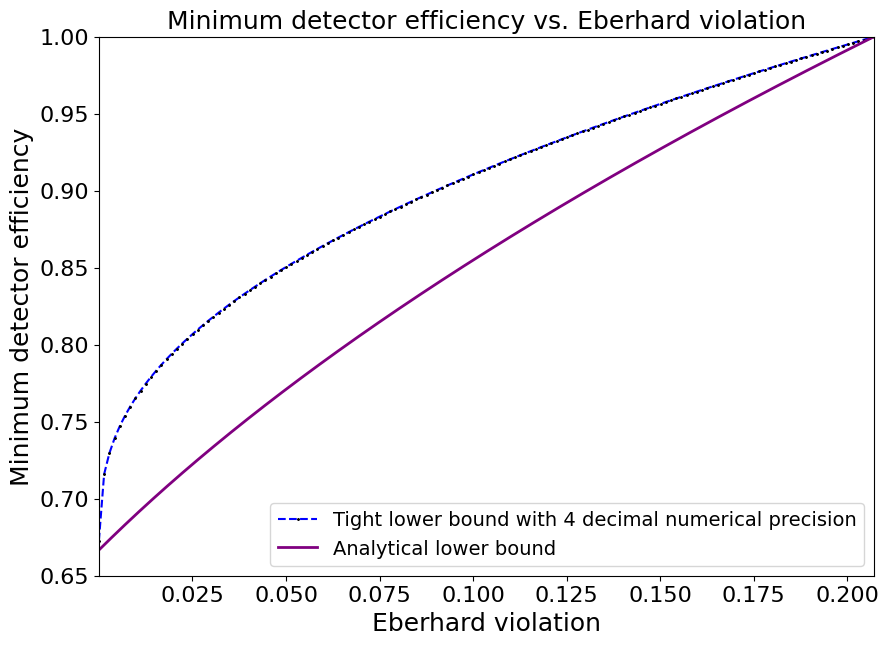}
    \caption{This figure represents two plots. The blue dotted curve reproduces the numerical plot in Fig.~\ref{numericalplot} of the minimum detector efficiency $\eta_{\mathcal{E}_{obs}}^{\min}$ (accurate up to $4$ decimals) over all quantum realizations for given Eberhard violation $\mathcal{E}_{obs}>0$. The violet curve represents the analytical lower bound on the detector efficiency $\eta_{ns}$ described in Eq.$~$(\ref{effns}), which is an explicit function of the variable $\mathcal{E}_{obs}\in (0, ~\frac{1}{2}(\sqrt{2}-1)]$.  }
    \label{analytical}
\end{figure}

While we do not pursue an improved analytical bound here, one could proceed in this direction by introducing additional linear constraints beyond the no-signaling conditions and the Tsirelson bound to obtain a tighter approximation to the quantum set. To elaborate, the distribution in~\eqref{pns} is not quantum realizable, as evidenced by its super-quantum resistance to detector inefficiency in Figure~\ref{analytical}: no quantum distribution can outperform the global optimum over NPA-realizable distributions. Therefore, one could introduce new inequalities, e.g.~the ``tilted'' CHSH inequality in~\cite{AcinMassarPironio}, to rule out the distribution in~\eqref{pns} and pursue an updated version of Proposition~\ref{prop1} in which the most robust distribution is less robust than~\eqref{pns} and yields a higher curve than~\eqref{effns}. Such an approach parallels the development of refined polytope approximations to the quantum set in~\cite{Bierhorst_2020}, recently improved upon in a work~\cite{Jee2025} where an Algorithm, which uses semi-definite programming to find the best Bell inequalities to rule out undesired super-quantum behaviors, could be adapted to rule out~\eqref{pns}. However, further development along these lines will presumably lead to more complicated formulas of increased complexity compared to~\eqref{effns}, possibly lacking a closed analytic form and/or requiring piece-wise definition over different ranges of $\mathcal{E}_{obs}$---a result of unclear value in consideration of the numerically tight bounds that can be obtained with the methods described in Section \ref{bigsec3}. 
\section{Conclusion}
Our study determined, with an accuracy up to $4$ decimal places, the minimum detector efficiency required in a symmetric setup to observe a desired Eberhard violation. 
We obtained such numerical accuracy through the agreement between 
a numerical upper bound on the minimum efficiency  found through direct search of quantum realizations, and a lower bound obtained with the second level of the NPA hierarchy. 
We also demonstrated that the method can be generalized to incorporate a nonzero dark count rate, and outlined an approach to further generalizations of linear noise models that encompass asymmetries of efficiencies between different parties and/or different settings.


 The above methods could not yield an analytical closed-form expression of the minimum efficiency over quantum realizations. To address this, we adopted an analytical approach based on the family of no-signaling behaviors that satisfy the Tsirelson bound. Although such set of behaviors forms a looser superset of the set of quantum realizations compared to the NPA hierarchy, these behaviors are analytically easier to characterize. We thereby derived a closed form expression of the minimum detector efficiency, which turns out to be a simple and monotonically increasing function of the Eberhard violation. However, we found that the analytical lower bound is noticeably lower than the numerically obtained, tight lower bound on the efficiency.

In Ref.~\cite{PhysRevLett.104.060401} it was shown that the minimum required detection efficiency in a symmetric Bell test involving two parties, with each party performing four two-outcome measurements can be reduced to $61.8\%$ with the violation of the so-called $I_{4422}$ Bell inequality \cite{BRUNNER20083162}. One may attempt to adapt our method for such inequality violation, 
which could allow device-independent certification of detector efficiency 
below 2/3, the lower limit of our current protocol. 
One may also go beyond two parties, where the violation of certain inequalities, such as for example the Mermin inequality \cite{PhysRevLett.65.1838} can demonstrate 
nonlocality in larger networks of parties. In light of the current increasing interest in quantum networks \cite{RevModPhys.95.045006}, this could be a fruitful direction for certifying efficiencies and more general noise tolerances for larger networks of parties.


\section*{Acknowledgements}
The authors acknowledge helpful discussions with Michael Mazurek and Lynden K.~Shalm. This work was partially supported by NSF Award Nos.~$2328800$ and $2435378$.

%

\newpage
\onecolumngrid
\appendix
\section{\label{appA}Expressions of the elements of a quantum realization \texorpdfstring{$\mathbf{P}= \lbrace P(ab|xy) \rbrace$}{P1} for the (2,2,2) setting based on the parametrization in Eq.~\eqref{parametrization}} 
According to the parametrization given in Eq.$~$\eqref{parametrization} one can evaluate the sixteen elements of the behavior $\mathbf{P} = \lbrace P(ab|x~y) \rbrace$ for the (2,2,2) setting using Eq.$~$\eqref{setting_for_alice} - \eqref{proj} as follows:

\begin{align}
     &P(++|x~y)= \dfrac{1}{8}\left[ 2 + \cos{(2\theta - \theta_x)}+ \cos{(2\theta + \theta_x)}+ 2~ \cos{\theta_y}~(\cos{2\theta}+ \cos{\theta_x})+ 2~ \sin{\theta_x}~\sin{\theta_y}~\sin{(2\theta)}\right]\nonumber \\
    \nonumber \\
    &P(+0| x~y) =\sin^2{\left( \dfrac{\theta_x}{2}\right)} \cos^2{\left( \dfrac{\theta_{y}}{2}\right)} \sin^2{\theta} + \cos^2{\left( \dfrac{\theta_{x}}{2}\right)} \sin^2{\left( \dfrac{\theta_{y}}{2}\right)} \cos^2{\theta} -\dfrac{1}{4}~\sin{\theta_{x}}~ \sin{\theta_{y}}~ \sin{(2\theta)} \nonumber \\
    \nonumber \\
     &P(0+| x_1y_0)= \dfrac{1}{8} \left[ 2+ \cos{(2\theta + \theta_{y})}+ \cos{(2\theta - \theta_{y})} - 2 ~\cos{\theta_{x}}~(\cos{(2\theta)} + \cos{\theta_{y}}) - 2  ~\sin{\theta_{x}}~\sin{\theta_{y}}~\sin{(2\theta)}\right]. \nonumber \\
     \nonumber \\
     &P(00|x~y) = \dfrac{1}{4} \left[ 1 + \cos{\theta_x}~\cos{\theta_y} - \left( \cos{\theta_x} + \cos{\theta_y} \right)~\cos{(2~\theta)} + \sin{\theta_x}~\sin{\theta_y}~\sin{(2~\theta)}\right],
\end{align}
where $\theta_x \in \lbrace \theta_{a_0},~\theta_{a_1} \rbrace$ for Alice's two measurement configurations and $\theta_y \in \lbrace \theta_{b_0}, ~\theta_{b_1} \rbrace$ for Bob's configurations, where the five independent parameters lie in the domain: \[0\leq \theta \leq \dfrac{\pi}{2} \quad and \quad 0 \leq \theta_{a_0}, ~\theta_{a_1}, ~\theta_{b_0},~ \theta_{b_1}\leq 2~\pi. \]

\section{Equivalence between Eberhard inequality and CHSH inequality} \label{appB}
Consider the set $\mathcal{NS}$ of all no-signaling behaviors $\mathbf{P}$ in the (2,2,2) scenario, where every $\mathbf{P}= \lbrace P(ab|xy)\rbrace$ can be written as 
\begin{align}
    P(ab|xy) = \dfrac{1}{4} \left[ 1 + a ~\langle \hat{A}_x \rangle + b~ \langle \hat{B}_y \rangle + ab~ \langle \hat{A}_x \hat{B}_y \rangle  \right], \label{eq.1}
\end{align}
where $a,b \in \lbrace +1, -1 \rbrace$ represents the outcomes and $\langle \hat{A}_x\rangle , ~ \langle \hat{B}_y \rangle,~ \langle \hat{A}_x \hat{B}_y \rangle$ denotes the marginals and correlators of $\mathbf{P}$ for the setting $x\in \lbrace x_0, y_0 \rbrace$ and $y\in \lbrace y_0, y_1 \rbrace$ which are defined in Eq.$~$(\ref{corr}) and in Eq.$~$(\ref{marginal}). We can write the Eberhard quantity $\mathcal{E}$ as follows: 
\begin{align}
    \mathcal{E} = P(++|x_0y_0) - P(+0|x_0y_1) - P(0+|x_1y_0) - P(++|x_1y_1), \label{evapp}
\end{align}
where the four joint probability terms can be expressed from Eq.$~$(\ref{eq.1}) as 
\begin{align}
     P(++|x_0y_0) &= \dfrac{1}{4} \left[ 1 +  \langle \hat{A}_{x_0} \rangle +  \langle \hat{B}_{y_0} \rangle +  \langle \hat{A}_{x_0} \hat{B}_{y_0} \rangle  \right] \nonumber \\
     P(+0|x_0y_1) &= \dfrac{1}{4} \left[ 1 +  \langle \hat{A}_{x_0} \rangle -  \langle \hat{B}_{y_1} \rangle -  \langle \hat{A}_{x_0} \hat{B}_{y_1} \rangle  \right] \nonumber \\
     P(0+|x_1y_0) &= \dfrac{1}{4} \left[ 1 -  \langle \hat{A}_{x_1} \rangle +  \langle \hat{B}_{y_0} \rangle -  \langle \hat{A}_{x_1} \hat{B}_{y_0} \rangle  \right] \nonumber \\
    P(++|x_1y_1) &= \dfrac{1}{4} \left[ 1 +  \langle \hat{A}_{x_1} \rangle +  \langle \hat{B}_{y_1} \rangle +  \langle \hat{A}_{x_1} \hat{B}_{y_1} \rangle  \right], \label{expressions}
\end{align} 
where notice that the outcomes $``+"$ (click) can be considered as $``+1"$ outcome and $``0"$ (no click) can be considered as $``-1"$ as the Eberhard's construction lump together the $``-1"$ click event with the ``no click'' event.  Now if we just put all the expressions of Eq.$~$(\ref{expressions}) into Eq.$~$(\ref{evapp}), one obtains the following:
\begin{align}
    \mathcal{E} &= -\dfrac{1}{2} + \dfrac{1}{4} \left[  \langle \hat{A}_{x_0} \hat{B}_{y_0} \rangle   +  \langle \hat{A}_{x_0} \hat{B}_{y_1} \rangle +  \langle \hat{A}_{x_1} \hat{B}_{y_0} \rangle - \langle \hat{A}_{x_1} \hat{B}_{y_1} \rangle  \right] \nonumber \\
    & = \dfrac{\beta_{CHSH}}{4}  - \dfrac{1}{2},
\end{align}
where $\beta_{CHSH} = \langle \hat{A}_{x_0} \hat{B}_{y_0} \rangle   +  \langle \hat{A}_{x_0} \hat{B}_{y_1} \rangle +  \langle \hat{A}_{x_1} \hat{B}_{y_0} \rangle - \langle \hat{A}_{x_1} \hat{B}_{y_1} \rangle $ is the CHSH quantity defined in Eq.$~$(\ref{CHSH}). 

\section{Technical aspects of the numerical optimization in Eq.~\eqref{numopt} and its equivalence with Algorithm~\ref{alg1}} \label{appC}
For any quantum realization $\mathbf{P}= \lbrace P(ab|xy) \rbrace$, we know from Eq.$~$(\ref{Ebquant}) that the quantities $\mathcal{K}$ and $\mathcal{L}$ are linear in terms of the elements of $\mathbf{P}$. Now, for a given amount of positive Eberhard violation $\mathcal{E}_{obs}>0$, we define a set of efficiency values as 
\begin{align}
\mathcal{S} = \left \lbrace \eta > 0 \mid \exists~ \mathbf{P} \in \mathcal{Q},~\mathcal{K}~\eta^2 - \mathcal{L}~\eta \geq \mathcal{E}_{obs}\right \rbrace. \label{feasible}
\end{align}
We show that the set $\mathcal{S}$ is upper closed, which means that if $\eta_0 \in \mathcal{S}$, then any $\eta \geq \eta_0$ is also a member of the set $\mathcal{S}$. 

Let $\mathbf{P}$ be any quantum realization, $\mathcal{E}_{obs}$ be a nonzero violation, and suppose there exists an efficiency $\eta_0 \in \mathcal{S}$ for which 
\begin{align}
\mathcal{K}~\eta_0^2 - \mathcal{L}~\eta_0 = \mathcal{E}_{obs} + \delta \label{assumption}
\end{align}
where $\delta>0$. Then for the same behavior $\mathbf{P}$, we can show 
\begin{align}
    \mathcal{K}~\eta^2 - \mathcal{L}~\eta \geq \mathcal{E}_{obs} + \delta \quad \quad \forall \eta \geq \eta_0.
\end{align}
Let $\Delta$ denote the difference between $ \mathcal{K}~\eta^2 - \mathcal{L}~\eta$ and $ \mathcal{K}~\eta_0^2 - \mathcal{L}~\eta_0$, then we can write the following: 
\begin{align}
    \Delta &= (\eta^2 - \eta_0^2)~\mathcal{K} - (\eta - \eta_0)~ \mathcal{L} \nonumber \\
    & = (\eta - \eta_0) \left[ (\eta + \eta_0)~\mathcal{K} - \mathcal{L}\right]. \nonumber
\end{align}
Since $\eta \geq \eta_0$, and also $\mathcal K$ is positive (which follows from \eqref{assumption} after noting the nonnegativity of $\eta_0$, $\mathcal E_{obs}$, and the sum of probabilities $\mathcal L$), the quantity $\Delta$ is non-negative if the following is true: 
\[
(\eta + \eta_0)~\mathcal{K} - \mathcal{L} \geq 0, \quad \textnormal{or} \quad \eta + \eta_0 \geq \dfrac{\mathcal{L}}{\mathcal{K}}. 
\]
The above holds because from our assumption in Eq.$~$(\ref{assumption}), we know that since $\mathcal{E}_{obs}>0 $,
\[
\eta_0 > \dfrac{\mathcal{L}}{\mathcal{K}}
\]
and this always implies that 
\[
\eta + \eta _0 \geq 2\eta_0 >  \dfrac{2~\mathcal{L}}{\mathcal{K}}\ge\dfrac{\mathcal L}{\mathcal K} 
\]
which proves $\mathcal{S}$ is upper closed. 

Now, since the set $\mathcal{S}$ is upper closed, there should exist an efficiency value such that all $\eta > \eta_{min}$ are in $\mathcal S$ and any $\eta < \eta_{min}$ cannot be a member of $\mathcal{S}$. 
So with the iterative bisection method mentioned in the Algorithm~\ref{alg1}, the computer searches for a lower bound of the set $\mathcal{S}$ while changing $\eta$ value at every iteration based on whether the previous $\eta$ was a member of the set $\mathcal{S}$ or not. However, it is not guaranteed that the search finds the true minimum efficiency $\eta_{min}$. This is because the search that is based on an optimization does not fit into a category of optimization problem for which there are solvers known to provably converge to a global optimum (such as linear, or convex).

\vskip 0.2cm

We now turn to another technical point. Note that for a given violation $\mathcal{E}_{obs}>0$, the numerical minimization problem in Eq.$~$(\ref{numopt}) requires the equality constraint $\mathcal{K}~\eta^2 - \mathcal{L}~\eta = \mathcal{E}_{obs}$ to be satisfied for a quantum realization. Whereas, Algorithm~\ref{alg1} requires the inequality constraint $\mathcal{K}~\eta^2 - \mathcal{L}~\eta \geq \mathcal{E}_{obs}$ to be satisfied, which determines whether $\eta\in \mathcal{S}$ for a given $\mathcal{E}_{obs}>0$. We now show that these two  requirements are equivalent in the sense that, if $\mathcal{K}~\eta^2 - \mathcal{L}~\eta > \mathcal{E}_{obs}$ is satisfied for some quantum realization, then there exists some other quantum realization satisfying an equality condition with $\mathcal{E}_{obs}$. 
So let us assume that for $\eta = \eta_0$, there is a quantum realization $\mathbf{P}\in \mathcal{Q}$, which satisfies the following equality condition, say for example,
\begin{align}
    \mathcal{K}~\eta_0^2 - \mathcal{L}~\eta_0 = \mathcal{E}_{obs} + 0.001. \label{feasible1}
\end{align}
Then, using the convex property of the quantum set $\mathcal{Q}$, it can be shown that there exists some other quantum realization $\mathbf{P}'$ and some efficiency value $\eta$ for which the following equality, 
\begin{align}
     \mathcal{K}'~\eta^2 - \mathcal{L}'~\eta = \mathcal{E}_{obs} 
\end{align}
holds, where $\mathcal{K}', ~\mathcal{L}'$ linearly depend on the elements of $\mathbf{P}'$ as given in Eq.$~$(\ref{Ebquant}). The explanation is given below: 
\vskip 0.2cm 
There exists a local deterministic behavior $\mathbf{P}_D$ with the elements $P_D(00|xy)= 1$ for all $x,y$ for which the following condition 
\begin{align}
    \mathcal{K}_D ~\eta^2 - \mathcal{L}_D ~\eta = 0 \label{LDequal}
\end{align}
is satisfied, where $\mathcal{K}_D, ~\mathcal{L}_D$ are linear combinations of the elements $\lbrace P_D(ab|xy) \rbrace$ as defined in Eq.$~$(\ref{Ebquant}). The reason the {\it r.h.s} of the above equation is zero is because $\mathcal{K}, ~\mathcal{L}$ in Eq.$~$(\ref{Ebquant}) do not include the terms $P(00|xy)$, and so $\mathcal{K}=\mathcal{L}=0$.  

Now we can find a quantum achievable behavior $\mathbf{P}'$ which can be expressed as a convex mixture of two behaviors as  
\begin{align}
    \mathbf{P}' = \lambda~ \mathbf{P} + (1 - \lambda) ~\mathbf{P}_D, \quad \quad \lambda \in (0,1), \nonumber
\end{align}
where the quantum behavior $\mathbf{P}$ satisfies the property given in Eq.$~$(\ref{feasible1}) 
and the the convex weight $\lambda$ can be defined as 
\[
\lambda = \dfrac{\mathcal{E}_{obs}}{\mathcal{K} ~\eta_0^2 - \mathcal{L} ~ \eta_0}.
\]
After this, using the expression of $\lambda$ and using Eq.$~$(\ref{LDequal}) we can express the function $\mathcal{K}' ~\eta_0^2 - \mathcal{L}' ~\eta_0 $ as follows: 
\begin{align}
  \mathcal{K}' ~\eta_0^2 - \mathcal{L}' ~\eta_0  &= \lambda ~\left( \mathcal{K}'' ~\eta_0^2 - \mathcal{L}''\eta_0\right)  \nonumber \\
  & + (1- \lambda) ~\left( \mathcal{K}_D ~\eta_0^2 - \mathcal{L}_D ~\eta_0\right) \nonumber \\
  & = \mathcal{E}_{obs}
\end{align}

\section{Finding an analytical lower bound on the detector efficiency for the entire range of quantum achievable Eberhard violation (\texorpdfstring{$0< \mathcal{E}_{obs}\leq(\sqrt{2}-1)/2$}{Eb})}\label{appD}


\vskip 0.3cm 
\paragraph*{{\bf Proposition 1.}}
  {\it Suppose $\mathcal{E}(\mathcal{P}_Q, ~\eta) = \mathcal{E}_{obs}>0$ holds for some no-signaling behavior $P_Q$ that violates the Eberhard inequality while satisfying the Tsireson bound. Then there exists a different no-signaling behavior $P_{ns}$ of the form \eqref{pns}, comprising a convex combination of $\sqrt 2 - 1$ weight on the the $PR_1$ box correlation and all remaining weight on (only) the three local deterministic correlations $D_4, D_8, D_{12}$, such that $\mathcal{E}(\mathcal{P}_{ns}, ~\eta_{ns}) = \mathcal{E}_{obs}$ is satisfied for some $\eta_{ns} \leq \eta $.} 

\begin{proof} Recalling \eqref{Pcorr}, any no-signaling behavior $P_Q$ violating the Eberhard inequality can be expressed as a convex combination
\begin{align*}
P_Q=p_{PR}~PR_1 + p_1~D_1 + p_4~D_4 + p_5~D_5 + p_8~D_8 + p_9 ~D_9 + p_{12}~D_{12}+ p_{14}~D_{14}+ p_{15}~D_{15},
\end{align*}
where $p_{PR}$ is the weight of the $PR$ box correlation and $p_1, p_4, p_8, p_9, p_{12}, p_{14}, p_{15}$ are weights on local deterministic (LD) distributions 
As $P_Q$ is also assumed to satisfy the Tsirelson bound (which it must if it is quantum realizeable), $p_{PR}\le \sqrt 2 - 1$ must hold as $p_{PR}/2$ is equal to the Eberhard violation, as explained at \eqref{EbpPR} of the main text.

Now, following \eqref{eb1} we can write
\begin{align}
    \mathcal{E}(P_{Q}, \eta) &= \left( \dfrac{3}{2}~p_{PR} + 2~(p_1+p_5+p_9)+ p_{14}+p_{15}\right)~\eta^2 - \left( p_{PR} + 2~(p_1+p_5+p_9)+ p_{14}+p_{15}\right)~\eta \nonumber \\
    & =  p_{PR}\left(\dfrac{3}{2}\eta^2 -\eta\right) + \alpha (\eta^2-\eta)\nonumber\\
   & \le p_{PR}\left(\dfrac{3}{2}\eta^2 -\eta\right)\nonumber\\
   & \le (\sqrt 2 - 1)\left(\dfrac{3}{2}\eta^2 -\eta\right)\label{e:dropalpha}.
\end{align}
The first inequality above holds because $\alpha =  2~(p_1+p_5+p_9)+ p_{14}+p_{15}$ must be a non-negative quantity and $f(\eta) = \eta^2-\eta$ is nonpositive for any value of $\eta \in [0,1]$; similarly, the second inequality holds by the nonnegativity of $g(\eta) = (3/2)\eta^2-\eta$ for $\eta \ge 2/3$, where we note $\eta$ does indeed exceed 2/3 by the assumption that $\mathcal{E}(\mathcal{P}_Q, ~\eta) = \mathcal{E}_{obs}$ is positive. If we take $P_{ns}$ to be a distribution as described in the statement of the theorem, $\mathcal{E}(P_{ns}, \eta)$ will be equal to $(\sqrt 2 - 1)\left[(3/2)\eta^2 -\eta\right]$, so by \eqref{e:dropalpha} we have
\begin{align*}
    \mathcal{E}(P_{Q}, \eta) \le \mathcal{E}(P_{ns}, \eta).
\end{align*}
To complete the theorem, note that if equality holds above, $\eta_{ns}$ can be taken to be $\eta$. If on the other hand the above inequality is strict, we will have 
$$
\mathcal{E}(P_{ns}, \eta) =  (\sqrt 2 - 1)\left(\dfrac{3}{2}\eta^2 -\eta\right) > \mathcal{E}_{obs}>0;
$$
i.e., $(\sqrt 2 - 1)g(\eta)>\mathcal E_{obs}$ for some value of $\eta \in (2/3, 1]$, and since $(\sqrt 2 - 1)g(2/3) = 0$, the intermediate value theorem guarantees the existence of a $n_{ns}\in (2/3,\eta)$ for which $\mathcal{E}(P_{ns}, \eta_{ns})=(\sqrt 2 - 1)g(\eta_{ns})=\mathcal E_{obs}$.
\end{proof}


\end{document}